%% file: homan.tex
\documentstyle[11pt,aaspp4,amssym,epsfig]{article}

\begin{document}

\date{\today}

\title{Detection and Measurement of Parsec-Scale Circular Polarization in Four AGN}

\author{D. C. Homan\altaffilmark{1} and J. F. C. Wardle\altaffilmark{2}}
\affil{Department of Physics, Brandeis University, Waltham, Massachusetts
 02454, USA}
\affil{}
\affil{\em{Appeared in the Astronomical Journal, Nov 1999, vol 118, pg 1942}}
\altaffiltext{1}{dch@quasar.astro.brandeis.edu}
\altaffiltext{2}{jfcw@quasar.astro.brandeis.edu}

\begin{abstract}
We present five epochs of 15 GHz VLBA observations of 13 AGN.  These observations
were specially calibrated to detect parsec-scale circular polarization and our calibration
techniques are discussed and analyzed in detail.  We obtained reliable
detections of parsec-scale circular polarization in the radio jets of four AGN: 
3C\,84, PKS 0528+134, 3C\,273, and 3C\,279.  For each of these objects our 
detections are at the level of $\sim 0.3-1\%$ local fractional circular 
polarization. Each individual detection has a significance in the range of
3 to 10 $\sigma$.  Our observations are consistent across 
multiple epochs (and different calibration techniques) in the sign and magnitude of the
circular polarization observed.  3C\,273 and 3C\,279 both undergo core outbursts 
during our observations and changes in the circular polarization of both sources 
are correlated with these outbursts.  In general, we observe the circular 
polarization to be nearly coincident with the strong VLBI cores of these objects; however,
in 3C\,84 the circular polarization is located a full milli-arcsecond south of
the source peak, and in the 1996.73 epoch of 3C\,273 the circular polarization
is predominately associated with the newly emerging jet component.  Our observations
support the theoretical conclusion that emission of 
circular polarization is a sensitive function of opacity, being strongest when 
the optical depth is near unity.  
Circular polarization may be produced as an intrinsic component 
of synchrotron radiation or by the Faraday conversion of linear to circular 
polarization.  Our single frequency observations do not easily distinguish between 
these possible mechanisms, 
but independent of mechanism, the remarkable consistency across epoch of 
the sign of the observed circular polarization suggests the existence of a 
long term, stable, uni-directional magnetic field.  Single dish 
observations of 3C\,273 and 3C\,279 at 8 GHz by Hodge and Aller suggest that 
this stability may persist for decades in our frame of observation.  
\end{abstract}


\section{Introduction}
\label{s:intro}

The detection and measurement of circular polarization in compact, 
extra-galactic radio sources has long been a difficult observational 
challenge.  The first reliable observations of integrated circular 
polarization were made by Gilbert and Conway\markcite{GC70} (1970) at 49 cm.  
Weiler and  de Pater\markcite{WdP83} (1983) cataloged and reviewed integrated measurements of 
circular polarization in a large number of extra-galactic radio sources. They found
no reliable measurements above 0.5\% with the majority of the observations 
$\lesssim$ 0.1\%.  
Jones\markcite{J88} (1988) used numerical simulations to 
model relativistic jets in compact radio sources and found that the 
{\em local} fractional circular polarization could be much higher than 0.5\%.

Circular polarization may be produced either as an intrinsic component of 
synchrotron radiation or by Faraday conversion of linear to
circular polarization\markcite {JOD77} (e.g. Jones \& O'Dell 1977).  Intrinsic  
circular polarization is produced directly by the radiating particles and serves as a 
probe of the magnetic field structure along the line of sight.   Faraday 
conversion, however, is a propagation effect, dominated by the lower energy relativistic 
particles in the jet, and therefore serves as a probe of the low energy end of the 
relativistic particle distribution. 

We present observations of parsec-scale circular polarization in four
compact, extra-galactic radio sources.  These observations were part of an 
ongoing program to monitor the structure and polarization of 12 rapidly 
variable compact radio sources, using the Very Long Baseline Array (VLBA) 
at $\lambda 1.3$ cm (22 GHz) and $\lambda 2$ cm (15 GHz).  
The circular polarization observations presented in this paper were 
made at 15 GHz.  Observations were made at intervals of two 
months, and the consistency between images made at different epochs is an 
important check on our results.  

We have detected circular polarization in 3C\,84, PKS 0528+134, 
3C\,273, and 3C\,279 at 15 GHz.  We find each of these sources to have
local circular polarization at levels of 0.3\% to 1\%.  These
observations are consistent across multiple epochs, and display changes 
linked with core outbursts in 3C\,273 and 3C\,279.

To make these observations we have devised three separate calibration and imaging
techniques, described in brief in section \ref{s:cp-cal}.  
Detailed discussion and analysis of these techniques and other
issues important to the measurement of circular polarization is postponed to 
the appendix.  The details of our experiments
and a brief description of our calibration procedures are provided in
section~\ref{s:details}.  Section~\ref{s:images} presents results
on all the sources in our sample and images of four of them.  We discuss the 
reliability of our calibration in \S{\ref{s:discuss-rel}}
and the physical interpretation of our observations in \S{\ref{s:discuss-phy}}.
Our conclusions are presented in section~\ref{s:conclusions}.

\section{Observational Details}
\label{s:details}

The observations presented here were part of a project to monitor closely a 
sample of 13 AGN at 15 and 22 GHz with the VLBA.\footnote{The National Radio
Astronomy Observatory is operated by Associated Universities, Inc., under cooperative
agreement with the National Science Foundation.}  The sources were
observed at two month intervals from January 1996 to December 1996, for 
a total of 6 epochs.  We present the 15 GHz circular polarization observations 
from the first 5 epochs. (Re-calibration for circular polarization of the 22 GHz 
companion observations is currently in progress.) The overall data quality of 
the sixth epoch is much poorer than the previous five making reliable 
detection of circular polarization difficult.  Table~\ref{t:epochs} lists 
the five epochs for which we present data in this paper.  
Table~\ref{t:sources} is a list of the 13 sources monitored 
with these observations.

\placetable{t:epochs}
\placetable{t:sources}

The observations were scheduled to maximize coverage in the u-v plane.  To accomplish
this goal, scan lengths were kept short (6.5 minutes per scan for the first two
epochs and 5.5 minutes per scan for the remaining epochs) and frequencies were
switched following every scan.  In addition, scans of neighboring sources were 
heavily interleaved at the expense of some additional slew time.  
Each source received approximately 45 minutes per frequency of observation 
time at each epoch. 

The data were recorded at each antenna using 1-bit sampling and were 
correlated at the VLBA correlator in Socorro, NM.  The correlator output 
contained 2 second integrations for 
all four cross-correlations (RR, RL, LR, LL), each with 4 IFs and 16
channels per IF.  The data were distributed on DAT tape to Brandeis
University where they were loaded into NRAO's Astronomical Imaging Processing
System (AIPS). 

Following a careful inspection of the data and supplied tables, the task
UVFLG read in the supplied flagging information and applied the 
initial edits.  The parallactic angles were then removed from the data
with the task CLCOR.  Prior to fringe fitting, the supplied pulse calibration
information was applied with the task PCCOR.\footnote{The only exception
is epoch 1996.05, where a manual phase calibration was used to align the phases across
IF. } Global fringe fitting was performed with the AIPS task FRING,
choosing a stable, central reference antenna such as Los Alamos (LA) or
Pie Town (PT).  The AIPS procedure CROSSPOL then removed the multi-band 
delay difference between the right and left hand systems of the
array. Initial amplitude calibration was from measured system temperatures
and atmospheric opacity (using the tasks ANTAB and APCAL).  A bandpass
correction was then determined with the task BPASS before averaging across
the channels within each IF.    

Following a short time-scale point source\footnote{After epoch-A, the source
model from a previous epoch was used in place of a point source model for the
initial phase coherence calibration.} 
phase calibration to increase
coherence, the data were averaged in time to 20 seconds and written out to
the Caltech VLBI program, DIFMAP.
The data were then edited in a
station based manner.  At this point, a copy of the data was written out
to disk; this data will be referred to as the edited, un-self-calibrated data.
Within DIFMAP, standard imaging and self-calibration techniques were
applied to obtain the best possible total intensity model.  This model
was then written out to AIPS and used to self-calibrate the edited,
un-self-calibrated data (using first a 20 second solution interval for phase
calibration followed by a 10 minute solution interval for amplitude
calibration). 

Following additional phase self-calibration passes in AIPS, 
the effects of feed leakage (D-terms) were removed from the
parallel and cross hand data.  This is crucial since the D-terms induce
non-closing errors in the RR and LL data that are not removed by
self-calibration and could mimic a circularly polarized signal.  On a single
baseline scan these errors are of order $D^2$ for an unpolarized source and $m_LD$ for a
polarized source, where $m_L$ is the fractional linear polarization. (See 
\S{\ref{s:dterms}} for a detailed analysis.)

The antenna D-terms were determined from observations of
the strong, compact source PKS 0528+134, using the program LPCAL\footnote{The
D-terms found by LPCAL are in the linear feed model.  Versions of AIPS
prior to the April 15, 1997 release require D-terms to be in the
ellipticity-orientation model if they are to be applied to all four
complex correlations. For this reason, it was necessary to translate the
D-terms to the ellipticity-orientation model prior to their application.}
in AIPS\markcite{LZD95} (Lepp\"{a}nen, Zensus, \& Diamond 1995).
The amplitudes of the D-terms ranged from approximately $1\%$ to
$5\%$ with values close to $2\%$ being most typical.  As described
in \S{\ref{s:dterms}}, our D-term calibration is accurate
enough to limit D-term related errors in our circular polarization images 
to less than $0.2\%$ of the corresponding linear polarization.

After D-term removal, a further amplitude self-calibration was
applied using the initial total intensity model from DIFMAP.  This is
important to remove any amplitude gain errors made
by the amplitude self-calibration prior to removal of the D-terms.  A
final round of phase self-calibration was performed before imaging in all four
Stokes parameters.  These images were made in the usual way and are
presented in Section~\ref{s:images}.

\section{Circular Polarization Calibration}
\label{s:cp-cal}

The calibration steps described above are similar to those used for
any VLBA observation designed to measure linear polarization 
\markcite{C93}\markcite{RWB94}(Cotton 1993; Roberts, Wardle, \& Brown 1994).  
For circular polarization, we must also carefully 
consider the relative calibration of 
the right ($R$) and left ($L$) complex antenna gains.  Appendix \ref{s:gain-cal} 
explores this issue in depth; 
here we briefly summarize our three separate techniques for calibrating the
 $R$ and $L$ complex antenna gains.

\subsection{Gain Transfer} 
\label{s:gt1}
In this technique, we initially make no assumption about the
presence or absence of circular polarization in our sources.  All self-calibration
rounds are performed with the rigorous assumption $(RR + LL)/2 = \tilde{I}_{model}$.
This assumption will calibrate only the average of $R$ and $L$ at a given antenna,
leaving the relative $R/L$ antenna gain ratio completely uncorrected.

At the end of the calibration only the $R/L$ gain ratios remain to be calibrated
before we can make circular polarization images.  If we know that some of our sources
have no circular polarization, we can calibrate this ratio on these sources by doing a
final self-calibration with the assumption: $RR = \tilde{I}_{model}$ and $LL = \tilde{I}_{model}$.  

We found these corrections (one per scan in amplitude and phase) on the subset of
sources which appeared to have no circular polarization (see \S{\ref{s:gain-trans}} for
discussion of how these sources were determined.).  The corrections were then 
averaged and smoothed on a 4 hour time scale before application to all sources in the
sample. The errors in this technique are dominated by the short-term (scan-to-scan)
$R/L$ gain fluctuations.  Analyzed in \S{\ref{s:gain-trans}}, these fluctuations
can limit the sensitivity of this technique to $\lesssim 0.15\%$ of the local
total intensity.  The errors listed in the tables of results were derived using
equation \ref{e:gain-errs} and we believe they provide a conservative estimate
of the $1 \sigma$ uncertainty.

Circular polarization images produced from data calibrated with this technique are 
referred to in the text as {\em gain transfer} images.  Making images in this way depends 
critically on the stability of the antenna gains.  It is a tribute to the outstanding 
performance of the VLBA that such images are now possible.

\subsection{Zero-V Self-calibration}
Another technique is to perform all self-calibration rounds with the assumption
that there is no real circular polarization (Stokes $V$) in the data.  In this calibration
scheme, separate complex gain corrections are found for the $R$ and $L$ hands
at each antenna by assuming $RR = \tilde{I}_{model}$ and $LL = \tilde{I}_{model}$.  
This assumption will try to reduce any circular polarization (real or spurious) 
as much as possible by adjusting the complex antenna gains.  We will refer to 
this calibration technique as {\em zero-V self-calibration}.

Real circular polarization is additive in the $RR$ and $LL$ correlations while the
complex gains are multiplicative.  Therefore, if a source has significant extended
structure, {\em zero-V self-cal} will not be able to completely remove
real circular polarization from the source.  In general, if the strong core
of a source has real circular polarization, this procedure will remove the circular
polarization from the strong core and transfer it (with reverse sign but the same
local fractional level) to the weaker extended structure.  If the circular
polarization is originally on the weaker extended structure (rather than the core),
{\em zero-V self-cal} cannot modify it much without inducing significant circular
polarization on the core.

\subsection{Phase-Only Mapping}
The previous two calibration techniques, {\em gain transfer} and {\em zero-V self-cal},
are subject to errors in the antenna amplitude gains.  The idea behind {\em phase-only}
mapping is to demonstrate that any circular polarization detected by these techniques
is also present (in a consistent manner) in the phases of the data alone.   If a source has
real circular polarization, then the $RR$ and $LL$ closure phases will be slightly different,
and this difference is preserved throughout the calibration process.  For simple sources, 
we can even use the {\em phase-only} image to predict the amplitudes of circular polarization
measured by the other techniques.

To construct a {\em phase-only} image, we set the amplitudes of the $RR$ and $LL$ 
correlations to unity and construct the $V$ visibilities from the resulting ``phase-only'' 
correlations.  Naturally, if the source is a simple point source or has no real
circular polarization, this technique will not detect any circular polarization. For
simple sources with extended structure and real circular polarization, {\em phase-only}
mapping produces circular polarization images that are anti-symmetric (to first order)
at the jet position around the location of the core.  

To construct {\em phase-only} images that are not dominated by short-term phase noise
fluctuations ($\sim 0.2^\circ$) between the $R$ and $L$ hands at each antenna, we found that 
the phases must be self-calibrated with the assumption $RR = \tilde{I}_{model}$ 
and $LL = \tilde{I}_{model}$.  However, as should be expected, the {\em phase-only}
images did not at all depend on how the amplitudes were corrected: calibrating the
amplitudes by assuming either $(RR+LL)/2 = \tilde{I}_{model}$ or $RR = \tilde{I}_{model}$, 
$LL = \tilde{I}_{model}$ produces identical results.
   
\section{The Observations}
\label{s:images}

Of the 13 observed sources, we have reliable detections of circular 
polarization on four: 3C\,84 (J0319+41), PKS 0528+134 (J0530+13), 3C\,273 (J1229+02), and
3C\,279 (J1256-05).  The 15 GHz observations of these four sources are 
presented individually.  The 15 GHz observations of the remaining 9 sources are
presented as a group.

\subsection{3C\,279 (J1256-05)} 

\begin{figure}
\epsfig{file=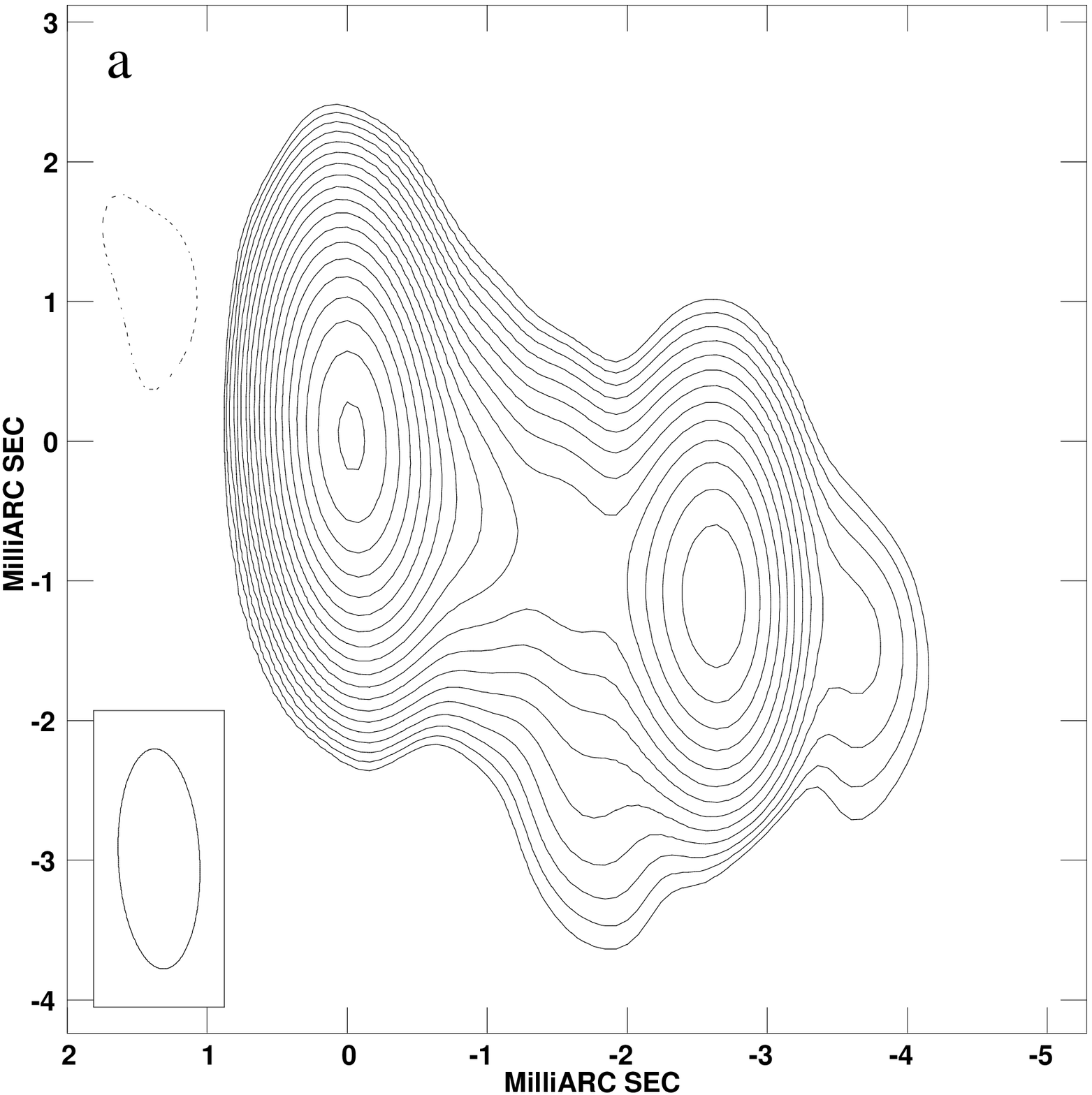,width=3.0in,angle=0}
\epsfig{file=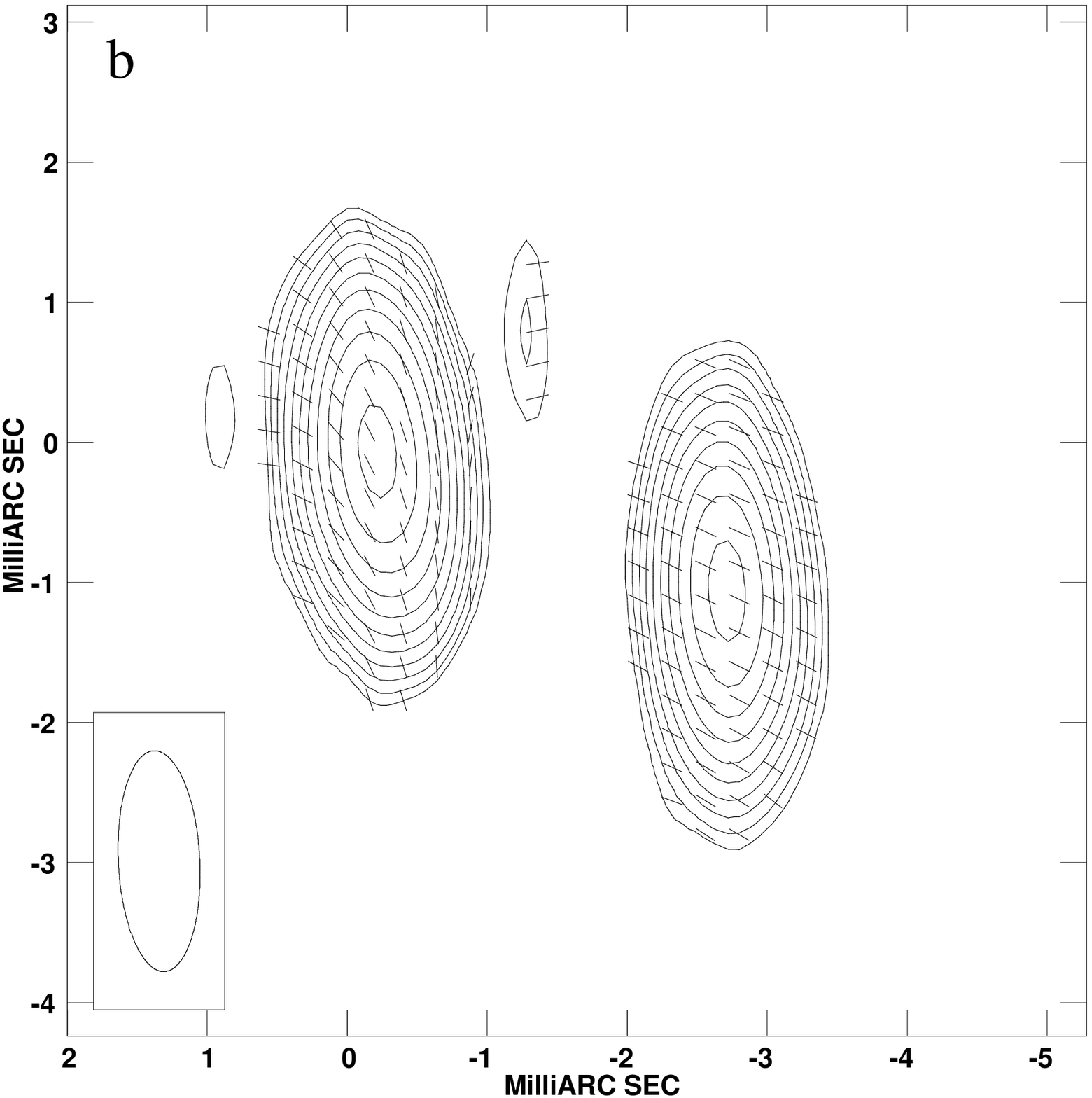,width=3.0in,angle=0} \\
\epsfig{file=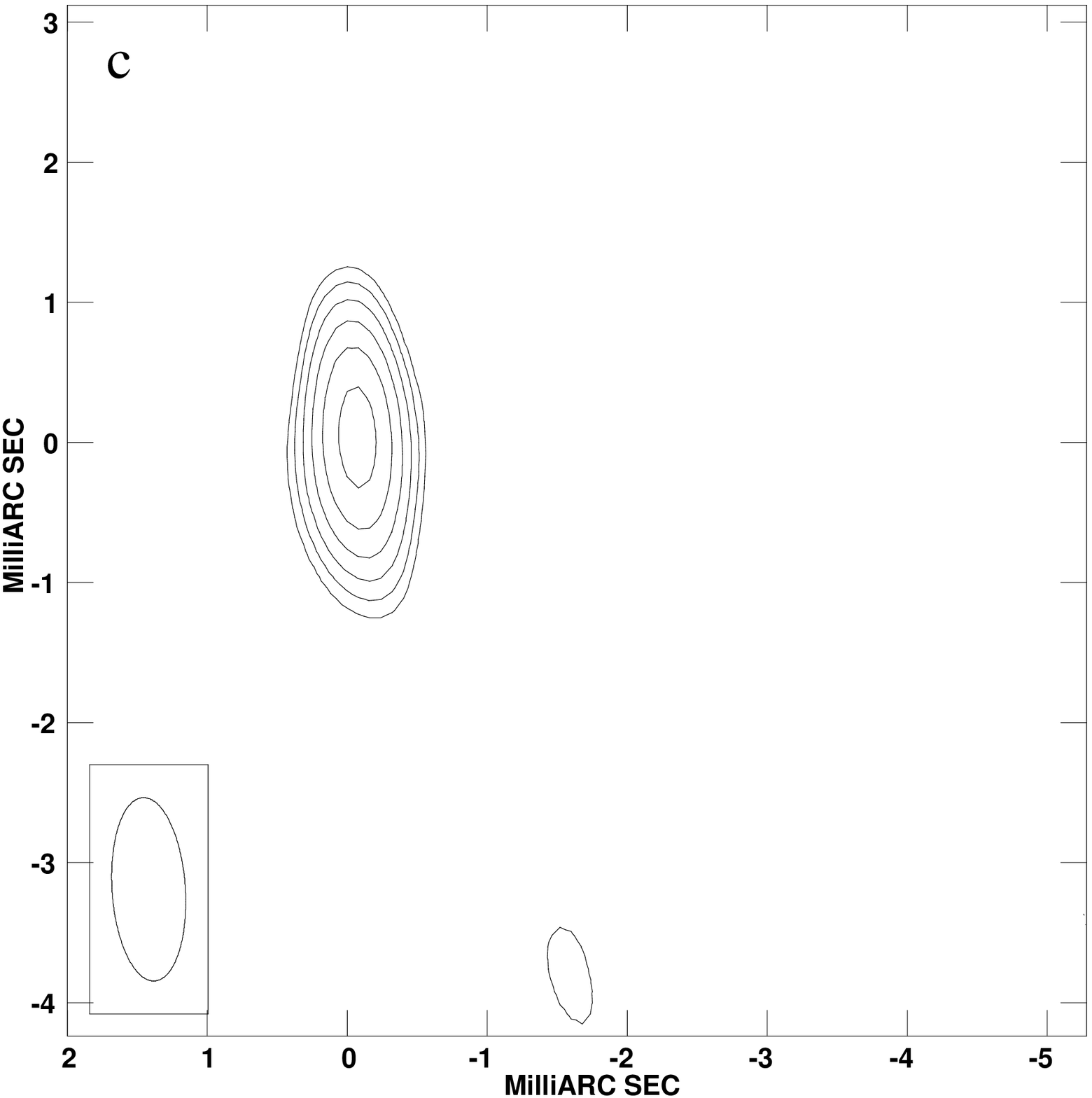,width=3.0in,angle=0}
\caption{\label{f:3c279-ipv}Naturally weighted images of 3C\,279, 
January 1996. a) Total intensity
$\sqrt{2}$ contours beginning at $0.02$ Jy/beam.  The map peak is
$10.92$ Jy/beam.  b) Linear polarization E-vectors superimposed on
polarization intensity with $\sqrt{2}$ contours beginning at $0.01$ Jy/beam.
The map peak is $0.36$ Jy/beam.  c) Circular polarization intensity
$\sqrt{2}$ contours beginning at $5$ mJy/beam.  
The map peak is $33$ mJy/beam.}
\end{figure}

\placefigure{f:3c279-ipv}

Images for the 1996.05 epoch of 3C\,279 in total intensity, linear polarization, 
and circular polarization are presented in Figure~\ref{f:3c279-ipv}.  
The circular polarization
image was produced with the {\em gain transfer} calibration technique 
(\S{\ref{s:gt1}}).  The images
are very similar to those from the later epochs (see Wardle et al.~\markcite{W98}(1998) for 
images from the 1996.57 epoch), although the core increases
in strength (as does the circular polarization) in the later epochs.  The images
display a strong, compact core with a jet extending to the southwest at 
position angle -113 degrees.   There is a bright knot in the jet at approximately
3 milli-arcseconds from the core. 

The core has modest fractional linear polarization ($m_L = 4.1\%$).  The observed
integrated circular polarization on the core is 36 mJy, corresponding to 
$+0.3\%$ local fractional circular polarization. While this measurement superficially
constitutes a 30$\sigma$ detection when compared to the RMS noise on the map, analysis
of the {\em gain transfer} technique in \S{\ref{s:gain-trans}} shows that such
measurements are limited by short time-scale gain fluctuations.  The associated errors
in the {\em gain transfer} circular polarization maps may be as large as $0.15\%$ 
of the corresponding total intensity.  Values obtained by the other calibration
techniques (described below), however, increase our confidence that our circular 
polarization measurement for epoch 1996.05 is accurate to a few milliJanskys.  

Detailed, u-v plane model-fitting
of the core region in total intensity and linear polarization (at both 15 and 22 GHz), 
reveals that the core consists of two closely spaced ($\sim 0.1$ mas) components.  
In Wardle et al.~\markcite{W98}(1998) we
argue that circular polarization is associated with the western core component (CW)
in the 1996.57 epoch.  Assuming this association holds for the other epochs, CW is
$+1.2\%$ circularly polarized in epoch 1996.05.  The core fluxes and polarizations 
for all five epochs are given in Table~\ref{t:3c279}.

\placetable{t:3c279}

\begin{figure}
\begin{center}
\epsfig{file=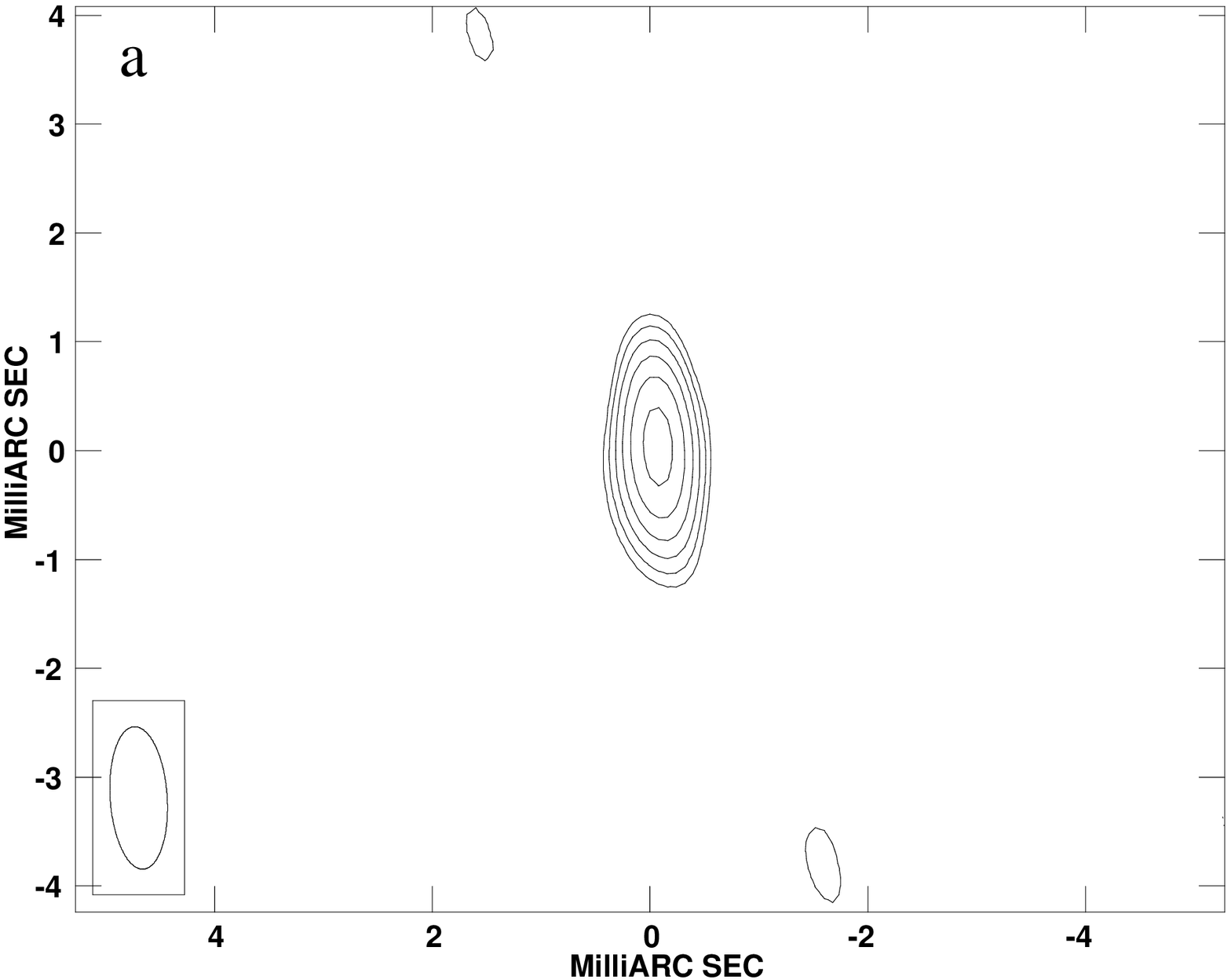,width=3.0in,angle=0} \\
\epsfig{file=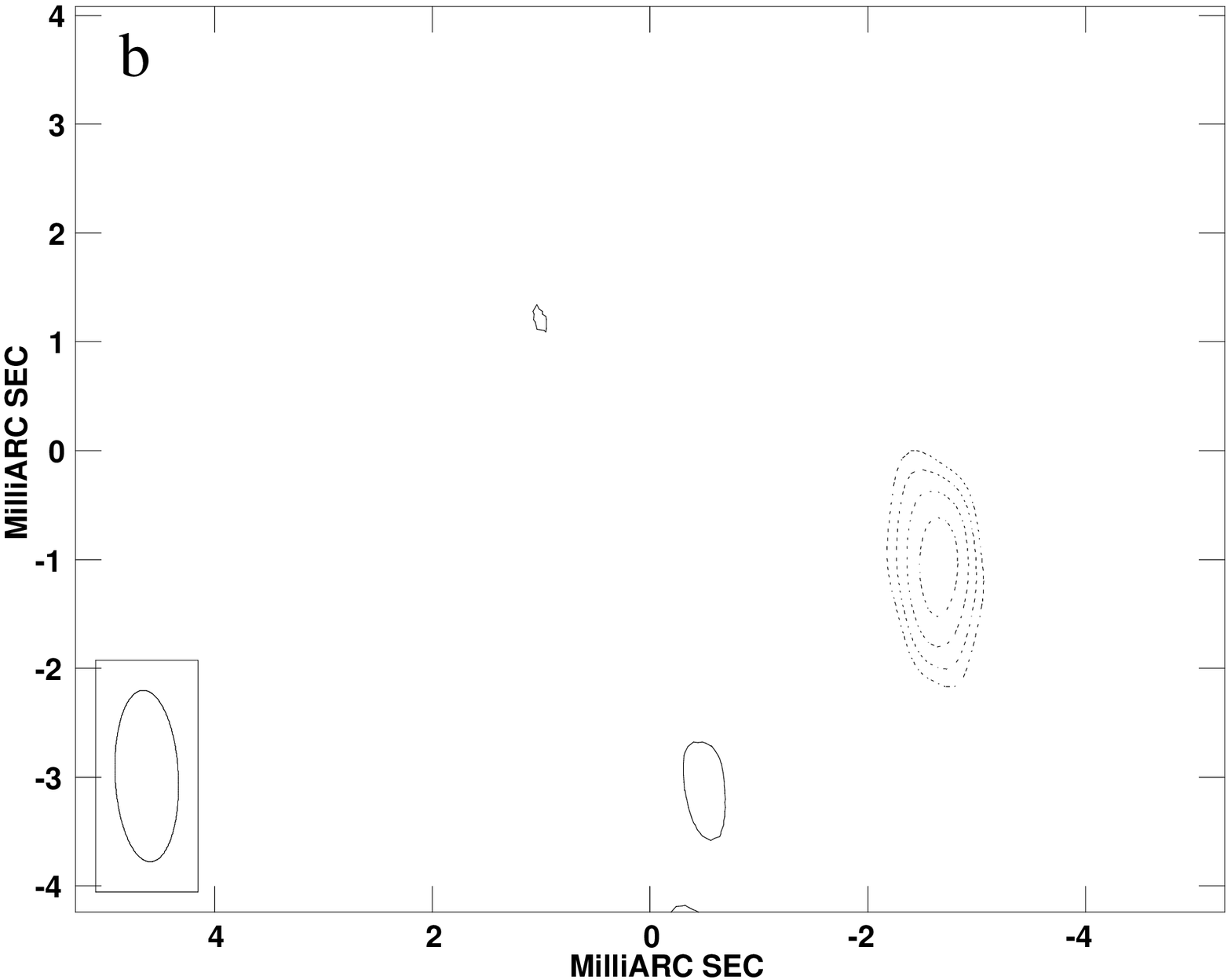,width=3.0in,angle=0} \\
\epsfig{file=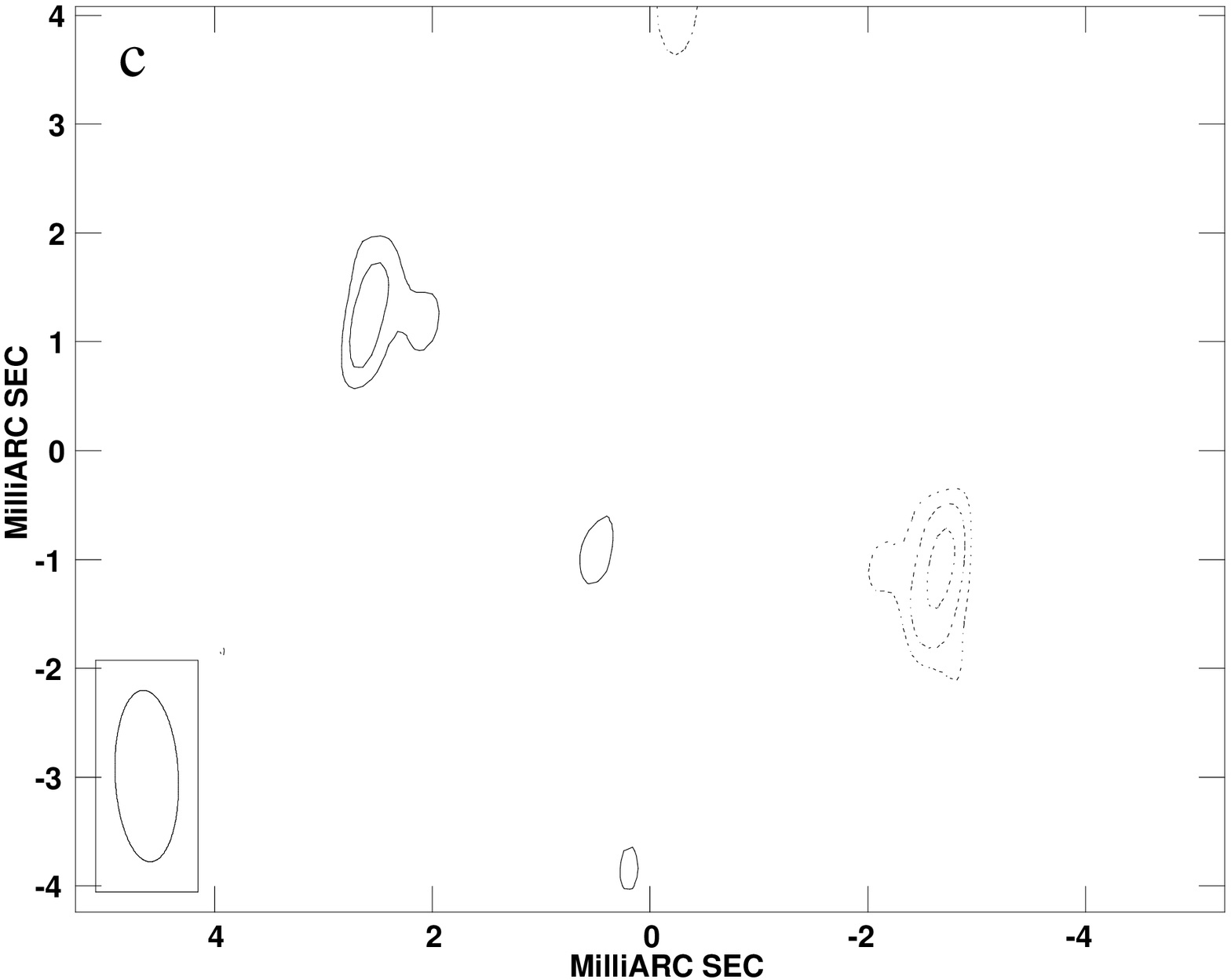,width=3.0in,angle=0}
\end{center}
\caption{\label{f:3c279-3v}Circular polarization images of 3C\,279, 
January 1996.  a) Gain
transfer calibration.  $\sqrt{2}$ contours begin at 5 mJy/beam.  The map
peak is 33 mJy/beam.  b) Zero-V self-calibration result.  $\sqrt{2}$
contours begin at +/$-$ 1.4 mJy/beam.  The map peak is $-$5 mJy/beam.  c)
Phase-only image.  $\sqrt{2}$ contours begin at 71 $\mu$Jy/beam.  The map peak
is $-$161 $\mu$Jy/beam.}
\end{figure}

3C\,279 provides our most robust circular polarization observations.
As a result of its relatively simple, (unequal) point double morphology on 
milli-arcsecond scales, all three of our calibration techniques give
readily interpretable and consistent results.  Figure~\ref{f:3c279-3v} displays 
the circular polarization images as produced by our three calibration 
techniques in epoch 1996.05.  

\placefigure{f:3c279-3v}

Figure~\ref{f:3c279-3v}a displays the image produced by our {\em gain transfer}
calibration procedure.  The image is the same as in Figure~\ref{f:3c279-ipv} except
that it has a different boundary for easy comparison to Figures~\ref{f:3c279-3v}b 
and~\ref{f:3c279-3v}c, produced by our other calibration procedures.  In epoch 1996.05,
the integrated core circular polarization measured by the {\em transfer
of gains} calibration is $+36$ ($\pm 18$) mJy.

Figure~\ref{f:3c279-3v}b displays the image produced by the
{\em zero-V self-cal}, calibration procedure.  The circular
polarization signal detected by this technique is located on the bright
knot in the jet at the $-0.3\%$ level.  As explained in \S{\ref{s:zerov}},
self-calibration assuming zero circular polarization will transfer circular
polarization from the strong core of an unequal point double (like
3C\,279) to the weaker jet component at the same local fractional level
but with reversed sign.  Assuming the circular polarization truly
lies on the core (as revealed by the {\em gain transfer} calibration 
technique), we can extrapolate the core circular polarization from
the observed jet circular polarization by modeling the source as a 
simple point double and using the relation, $V_1 \simeq 
-V_2(I_1/I_2)$ (see \S{\ref{s:zerov}}).  Applying this simple relation to
epoch 1996.05, gives an integrated core circular polarization of 
$+32$ ($+9$,$-3$) mJy.  

Figure~\ref{f:3c279-3v}c displays the image produced by our {\em phase-only} method.
The circular polarization detected by this technique is anti-symmetric
about the core with the jet knot location. The integrated intensities of
$-228$ $\mu$Jy for the jet and $+198$ $\mu$Jy for the anti-jet from the
phase-only data can be used to predict the core circular polarization,
using equation~\ref{e:phase-only} (\S{\ref{s:po2}})
 and assuming a simple point double model for 3C\,279.
Applying this relation to epoch 1996.05, gives an integrated
core circular polarization of $+27$ ($+7$,$-2$) mJy.

It is important to note that the errors associated with the {\em zero-V self-cal}
and {\em phase-only} measurements of circular polarization reflect not only
random noise but also the possible systematic loss of signal, described 
in \S{\ref{s:zerov}}, associated with assuming no circular polarization
at one or more steps in the self-calibration process.

\placetable{t:3c279-alt}
\placetable{t:3c279-2pt-ext}

\begin{figure}
\begin{center}
\epsfig{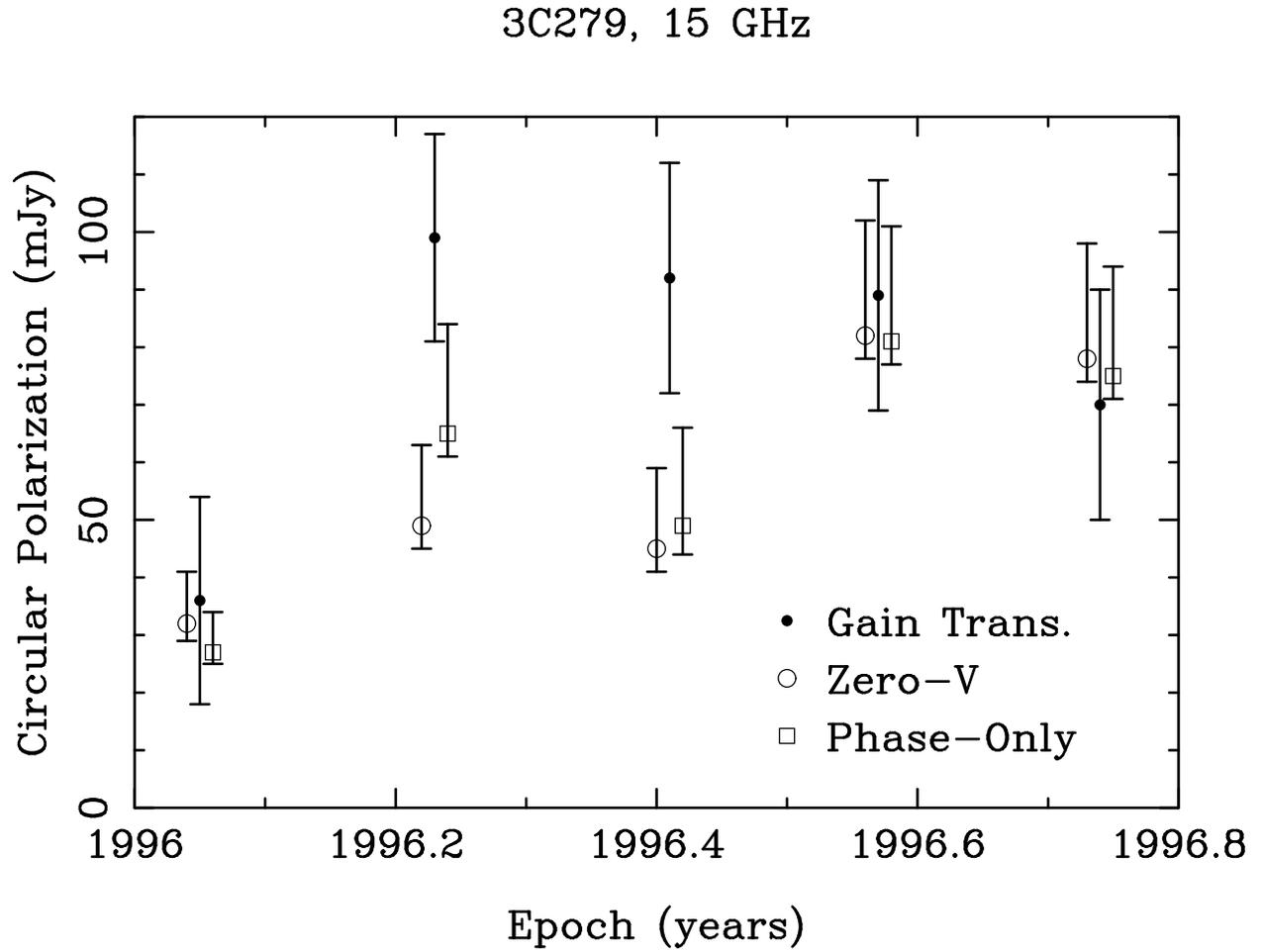}
\end{center}
\caption{\label{f:3c279-2pt-ext}
Circular polarization measurements of 3C\,279.  Results from all
three calibration techniques are plotted versus epoch.  Points for
an individual epoch are slightly displaced in time to avoid
confusing overlap of the error bars.}
\end{figure}

The {\em gain transfer}, {\em zero-V self-cal}, and {\em phase-only} maps
from the later epochs are very similar to those displayed in Figure~\ref{f:3c279-3v}
for epoch 1996.05.  Table~\ref{t:3c279-alt} displays the raw measurements of
circular polarization from the {\em zero-V self-cal} and {\em phase-only}
calibration techniques.
Table~\ref{t:3c279-2pt-ext} summarizes the results from applying a
simple point double model to these raw measurements and extrapolating 
the circular polarization of the core.  The extrapolations are compared
to the more direct {\em gain transfer} results.  Figure~\ref{f:3c279-2pt-ext}
compares the results from the three calibration techniques.

\placefigure{f:3c279-2pt-ext}

\subsection{3C\,84 (J0319+41)}

\begin{figure}
\begin{center}
\epsfig{file=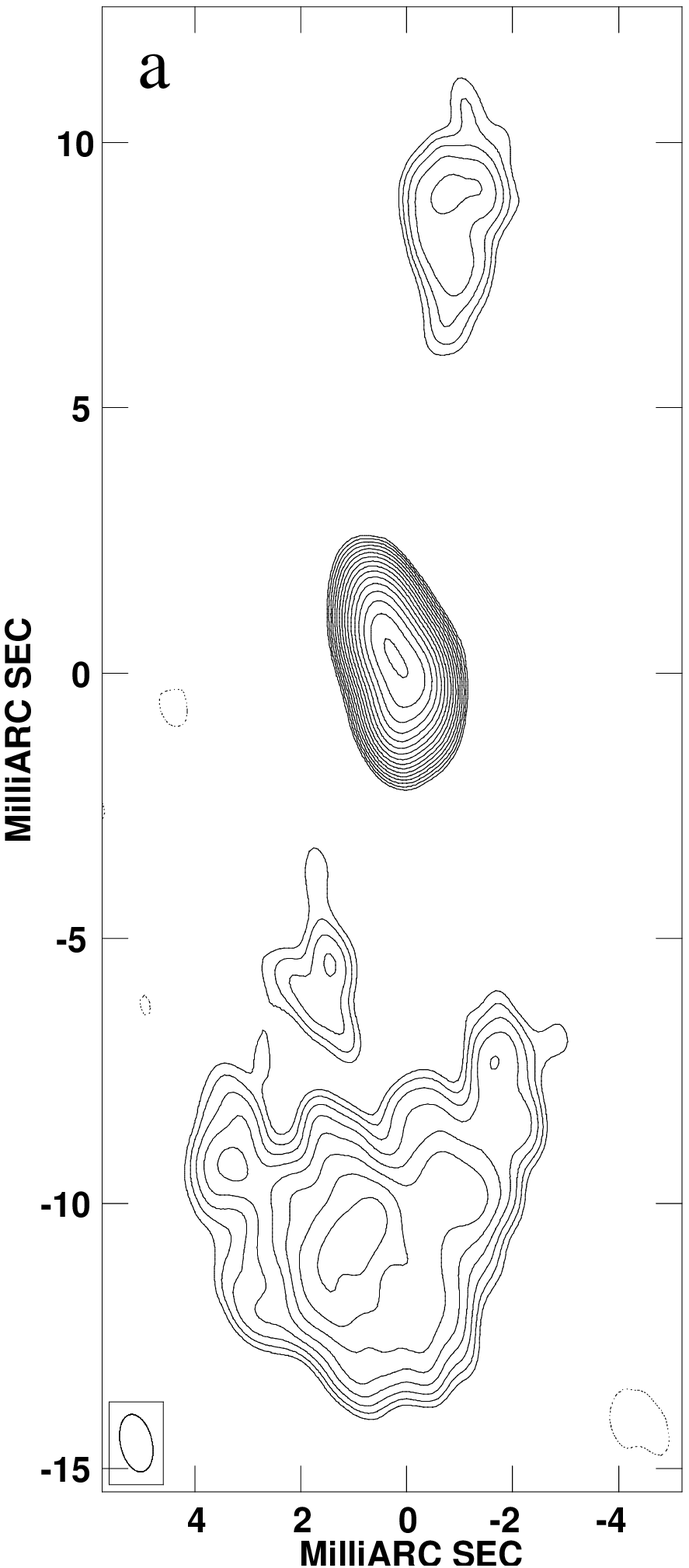,width=3.0in,angle=0}
\epsfig{file=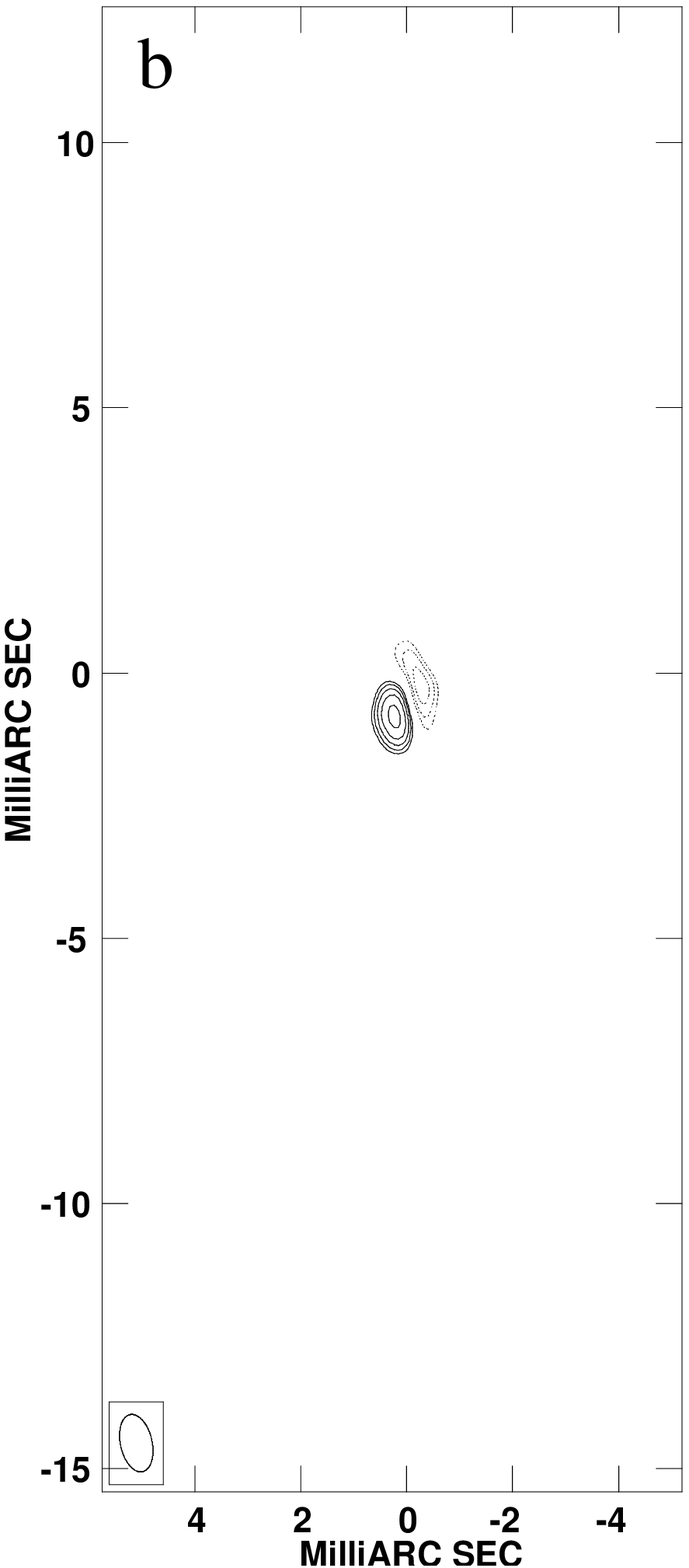,width=3.0in,angle=0}
\end{center}
\caption{\label{f:3c84-ipv}
Naturally weighted images of 3C\,84, September 1996.  a)
Total intensity $\sqrt{2}$ contours beginning at 0.020 Jy/beam.  The map
peak is 3.79 Jy/beam.  b) Circular polarization intensity map, produced with
the Zero-V self-cal technique, with $\sqrt{2}$ 
contours beginning at +/$-$ 4 mJy/beam.  The map peak is +18.1 mJy/beam.}
\end{figure}

\placefigure{f:3c84-ipv}

Figure~\ref{f:3c84-ipv} displays images for the 1996.74 epoch of 3C\,84 in both total
intensity and circular polarization structure.  3C\,84 has a strong core
which is elongated in the north-south direction.  The jet extends to the
south forming a diffuse lobe-like region about 10 mas south of the core.  
To the north we clearly detect the counter feature reported by Vermeulen,
Readhead, and Backer\markcite{VRB94} (1994) and Walker, Romney, and Benson 
\markcite{WRB94} (1994).  The
total intensity structure of 3C\,84 does not appear to change
significantly over the course of our observations.  The linear
polarization structure (not shown) of 3C\,84 is very weak.  Only the
``lobe'' shows significant linear polarization in a couple isolated
spots at the 1\% local level.

Figure~\ref{f:3c84-ipv}b is produced by {\em zero-V self-calibration} and clearly shows
the circular polarization structure to lie in the south-eastern part of 
the elongated core at approximately the $+1$\% ($\pm 0.1$\%) level.  A smaller, negative
piece of circular polarization is located just north and west of the
positive circular polarization. In 
\S{\ref{s:zerov}}, we show that {\em zero-V self-cal} does very little to modify 
the circular polarization in a source if it is well separated from the main
peak in total intensity.  Here the $V$ peak is only one beam-width south of the $I$ peak, and
we believe the negative component is an artifact resulting from the {\em zero-V self-calibration}
assumption. We were indeed able
to produce just such a negative signal by {\em zero-V self-cal} in simulated
data sets of 3C\,84. The images
produced by {\em zero-V self-cal } in the other epochs are essentially the
same as figure~\ref{f:3c84-ipv}b, each revealing about $+1$\% ($\pm 0.1$\%) local circular 
polarization in the same location.

\begin{figure}
\begin{center}
\epsfig{file=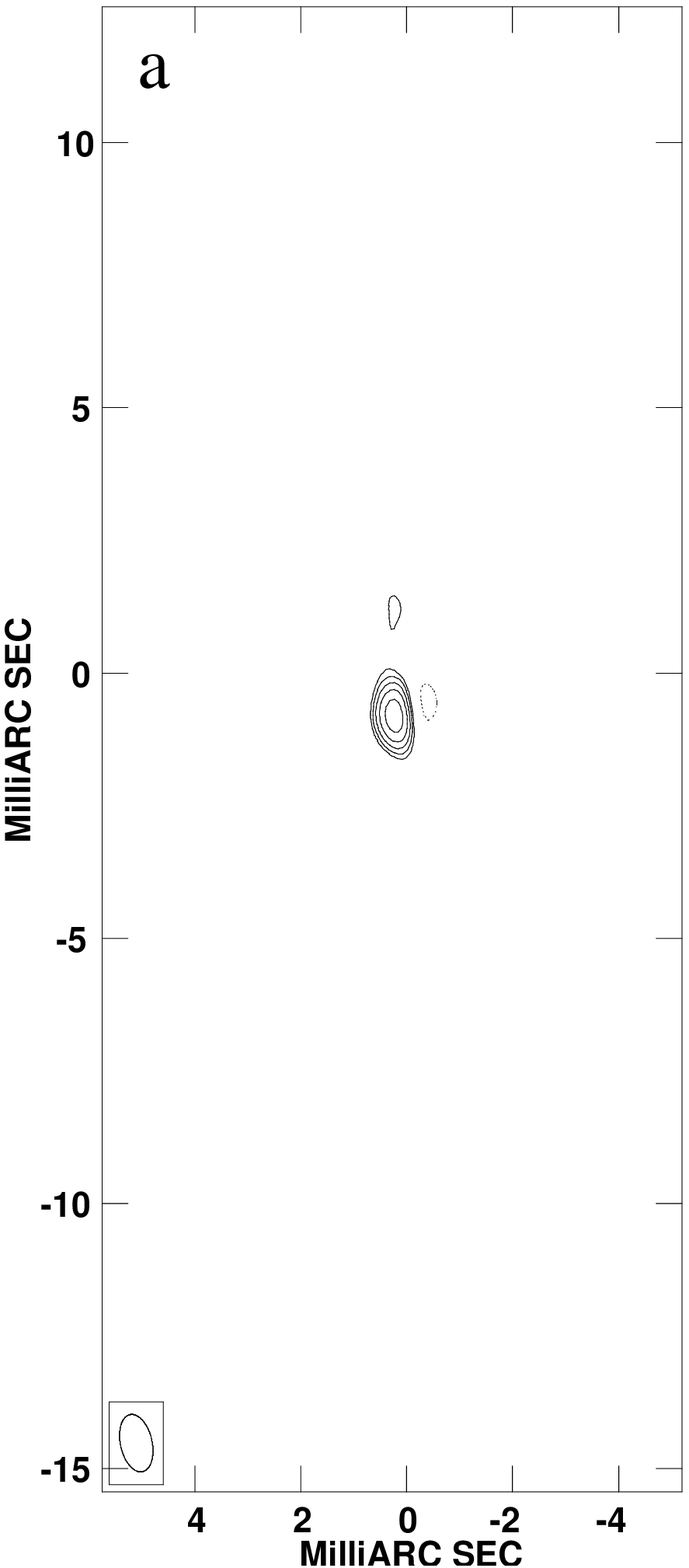,width=3.0in,angle=0}
\epsfig{file=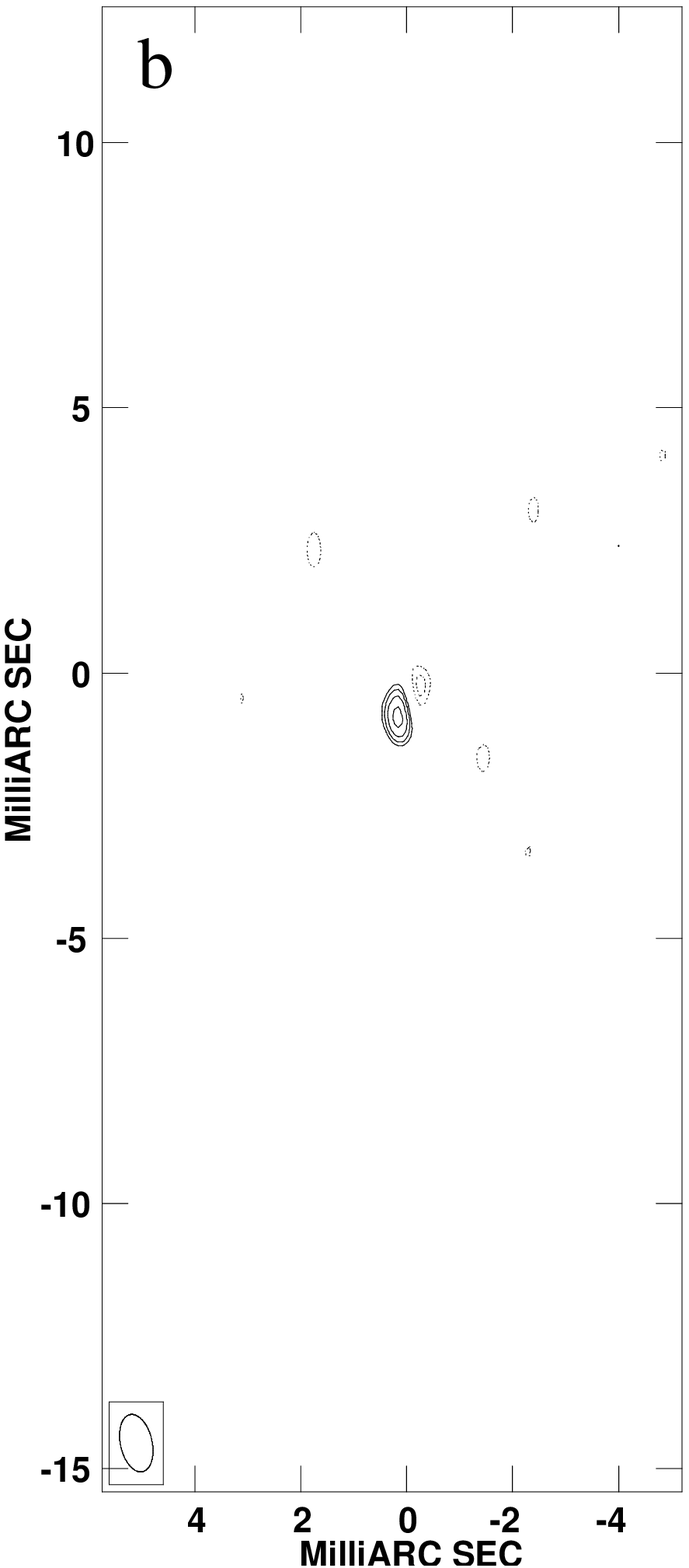,width=3.0in,angle=0}
\end{center}
\caption{\label{f:3c84-3v}
Circular polarization images of 3C\,84, May 1996.  a) Gain
transfer calibration.  $\sqrt{2}$ contours begin at +/$-$ 5.67 mJy/beam.
The map
peak is +28.3 mJy/beam.  b) Result of phase-only mapping.  $\sqrt{2}$
contours begin at +/$-$ 2 mJy/beam.  The map peak is +6.4 mJy/beam.}  
\end{figure}

The {\em gain transfer} and {\em phase-only} images are presented in
Figure~\ref{f:3c84-3v}.  They are both very consistent with the {\em zero-V self-cal}
result.  The {\em gain transfer} image (Figure~\ref{f:3c84-3v}a) shows the positive
circular polarization to be stronger (28 versus 18 mJy/beam) than
the {\em zero-V self-cal} image. The smaller, negative circular polarization
is also shown by the {\em gain transfer} image, although it is
much weaker and located more to the west rather than to the
north-west as indicated by the {\em zero-V self-cal} image.  

\placefigure{f:3c84-3v}

The other epochs show generally consistent {\em gain transfer} results, 
although the
quality of the {\em gain transfer} results vary from epoch to epoch.
The circular polarization signals measured by {\em gain transfer} 
are always in the same location (south and east of the map peak) and 
are usually within a couple of mJy of the {\em zero-V self-cal} 
result. 
The worst {\em gain transfer} result is the 1996.57 epoch which
has large pits and valleys in the circular polarization image, masking
the circular polarization signal clearly detected by the other 
two techniques.  

Figure~\ref{f:3c84-3v}b is the {\em phase-only} circular polarization image 
of 3C\,84 in the 1995.74 epoch.
The result is difficult to interpret directly due to the extreme complexity
of the source. However it is clear that this image is consistent with
the results produced by the other two techniques.  The {\em phase-only}
results from the other epochs are equally consistent.

\subsection{PKS 0528+134 (J0530+13)}

\begin{figure}
\epsfig{file=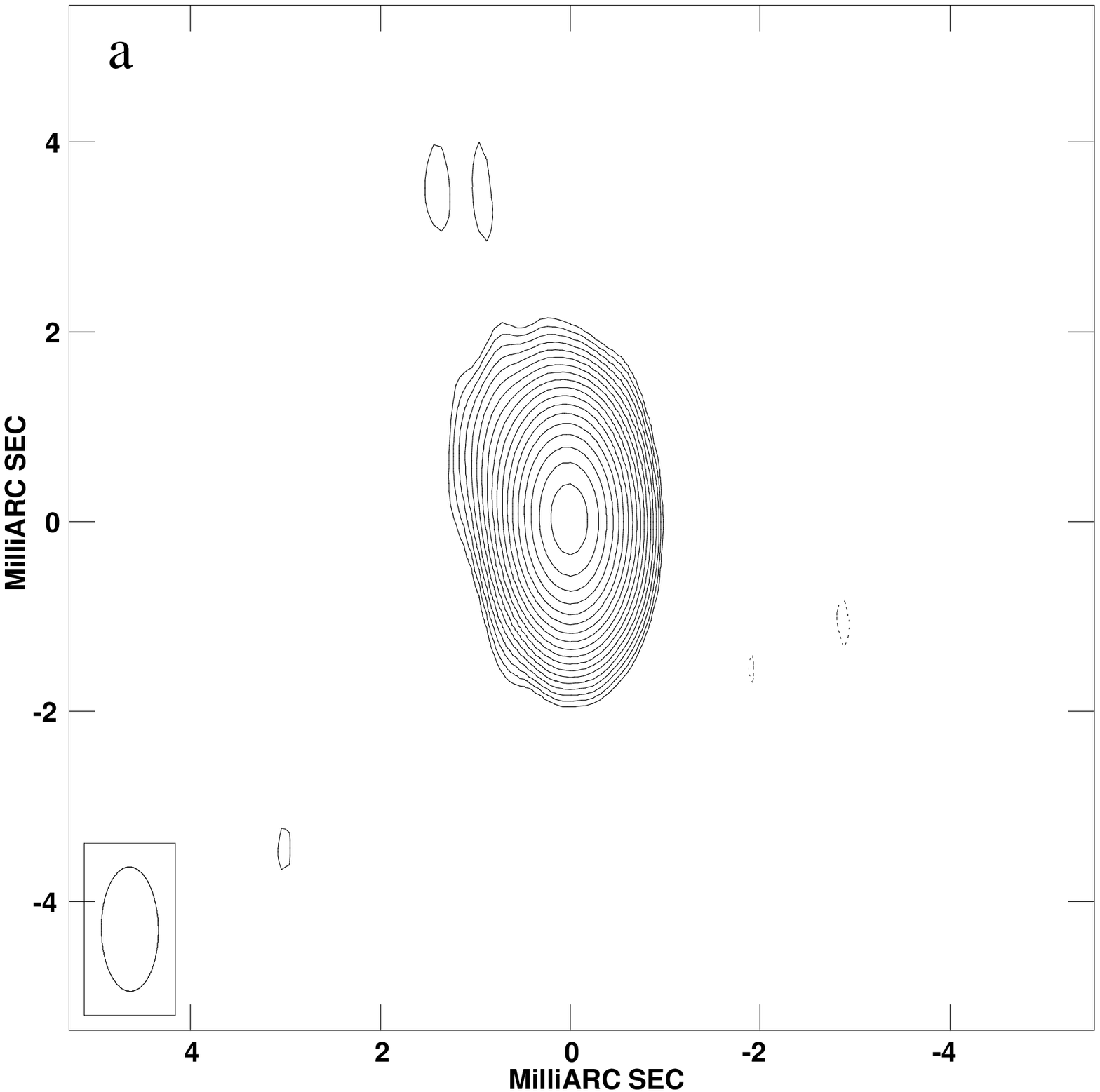,width=3.0in,angle=0}
\epsfig{file=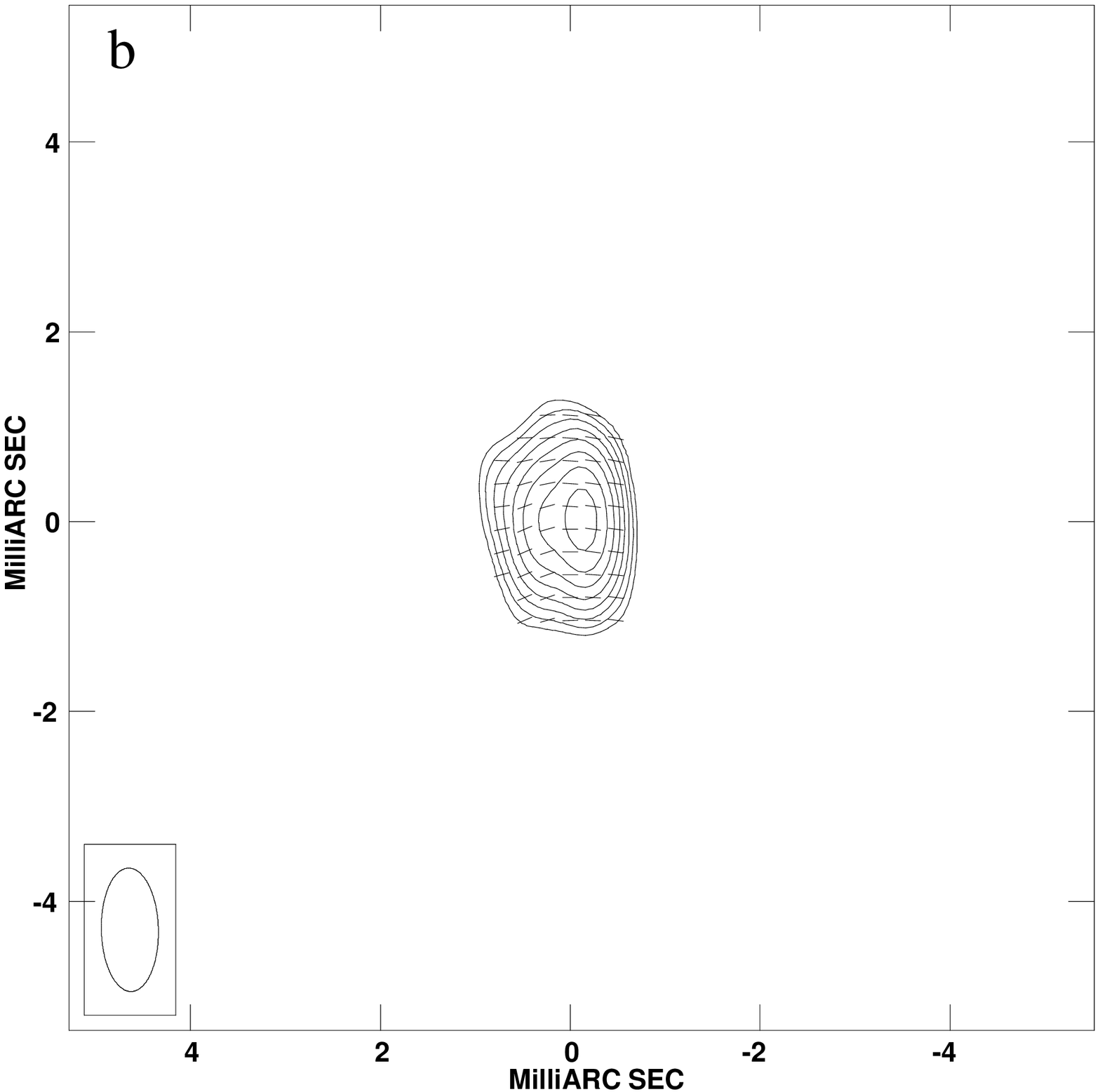,width=3.0in,angle=0} \\
\epsfig{file=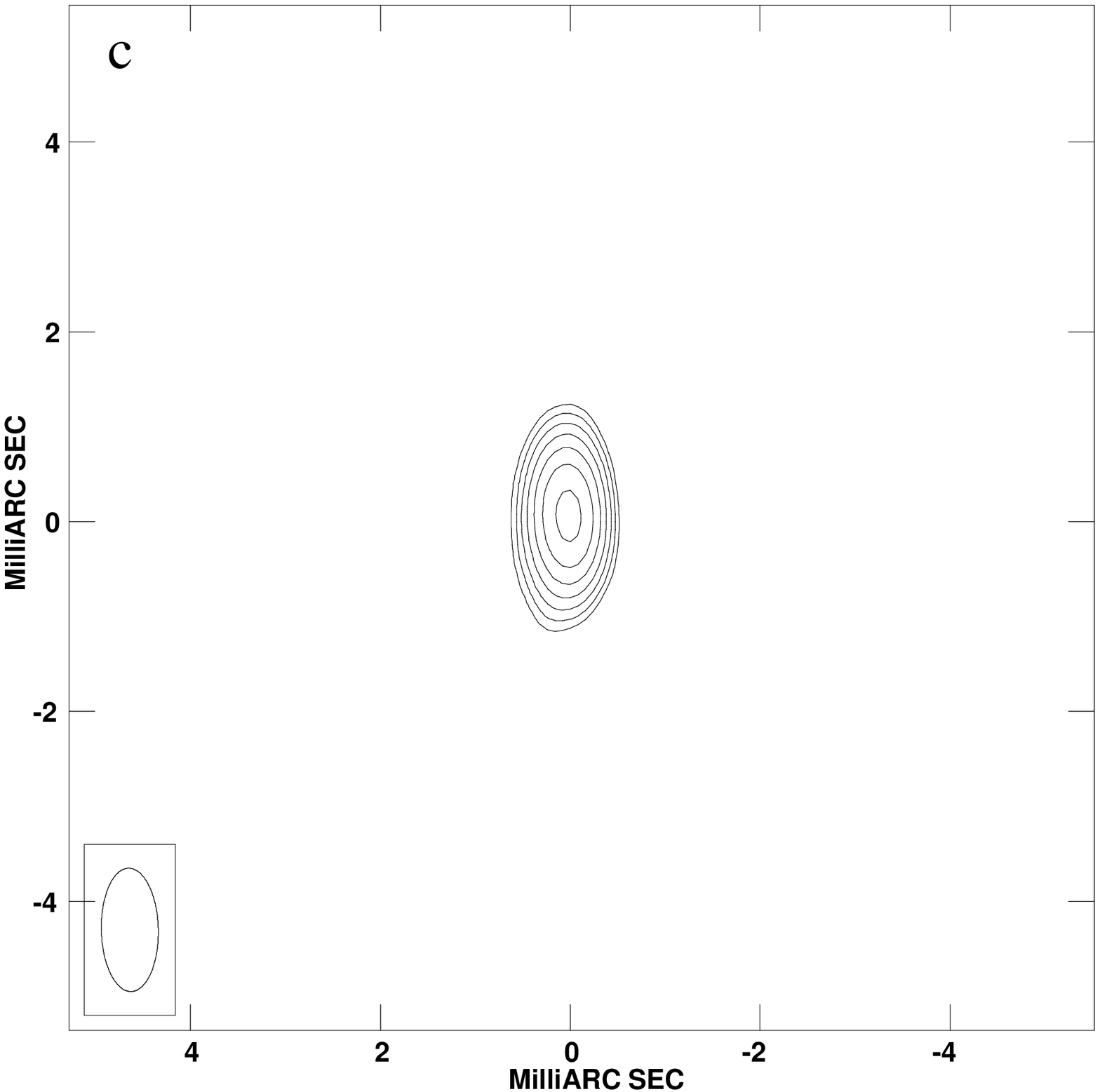,width=3.0in,angle=0}
\caption{\label{f:J0530-ipv}
Naturally weighted images of PKS 0528+134, July 1996.  a)
Total intensity $\sqrt{2}$ contours beginning at 0.015 Jy/beam.  The map
peak is 6.78 Jy/beam.  b) Linear polarization E-vectors superimposed on
polarization intensity with $\sqrt{2}$ contours beginning at 0.010 Jy/beam.
The map peak is 0.135 Jy/beam.  c) Circular polarization intensity
$\sqrt{2}$ contours beginning at 5 mJy/beam.  The map peak is 45
mJy/beam.}   
\end{figure}

\placefigure{f:J0530-ipv}

The total intensity, linear polarization, and 
circular polarization images of PKS 0528+134 from our 1996.57 epoch 
are presented in Figure~\ref{f:J0530-ipv}.  The total
intensity and circular polarization observations from the other epochs
are very similar to those presented.  The linear polarization, however,
changes considerably over the course of our observations.  PKS 0528+134 is 
a barely resolved point source with an extension to the NNE.  In tapered 
images we observe extended flux approximately 3 milli-arcseconds to the
NNE, suggesting that the parsec-scale jet extends in that direction.
The circular polarization (as revealed by the {\em gain transfer} 
calibration technique) lies directly on the core with an integrated
intensity of $+45$ mJy $\pm 10$ mJy ($m_c = +0.6\%$).  The circular
polarization maps for the other epochs are very similar, each revealing 
strong positive circular polarization: $+40$ to $+50$ mJy in three of
the other epochs and $+100$ mJy in one epoch.  

Our model-fitting of the total intensity data reveals the core to consist
of two closely spaced components, however the linear polarization
structure is fit quite poorly by this model.  We cannot associate
the circular polarization with either of these total intensity model
components with any confidence.  As a result, we will treat the core
as a single component with the associated linear and circular 
polarization. Table~\ref{t:J0530} contains the 
integrated measurements of total intensity,
linear polarization, and circular polarization for the core of PKS
0528+134 for each of the five epochs presented in this paper.

{\em Zero-V self-cal} and {\em phase-only imaging} both revealed no
detectable circular polarization on PKS 0528+134.  As explained in 
 \S{\ref{s:gain-cal}}, neither of these 
techniques would detect circular
polarization on what is essentially a point source. However, we are confident
that the circular polarization revealed in the {\em gain transfer} images is
real.

\placetable{t:J0530}

\subsection{3C\,273 (J1229+02)}

\begin{figure}
\epsfig{file=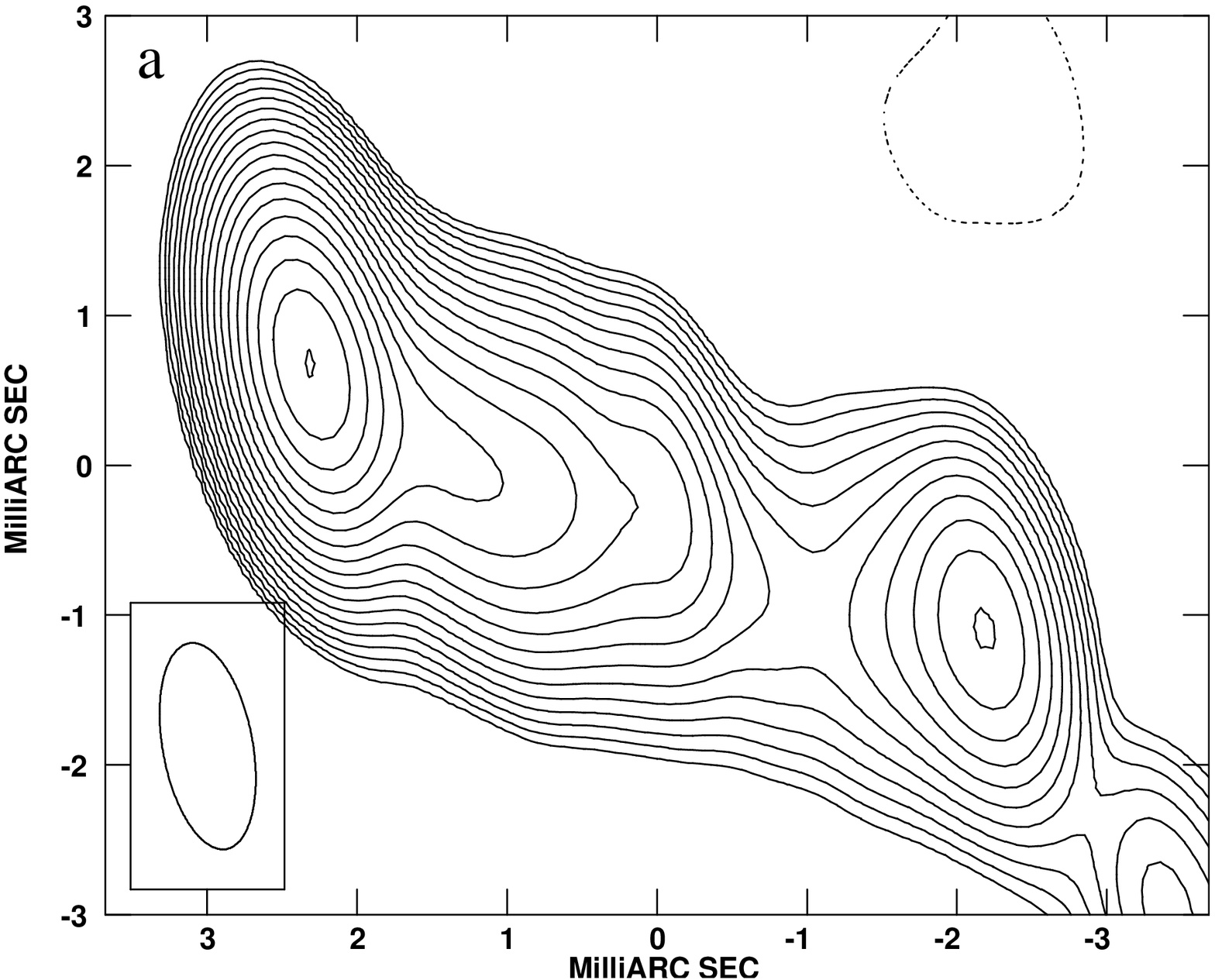,width=3.0in,angle=0} 
\epsfig{file=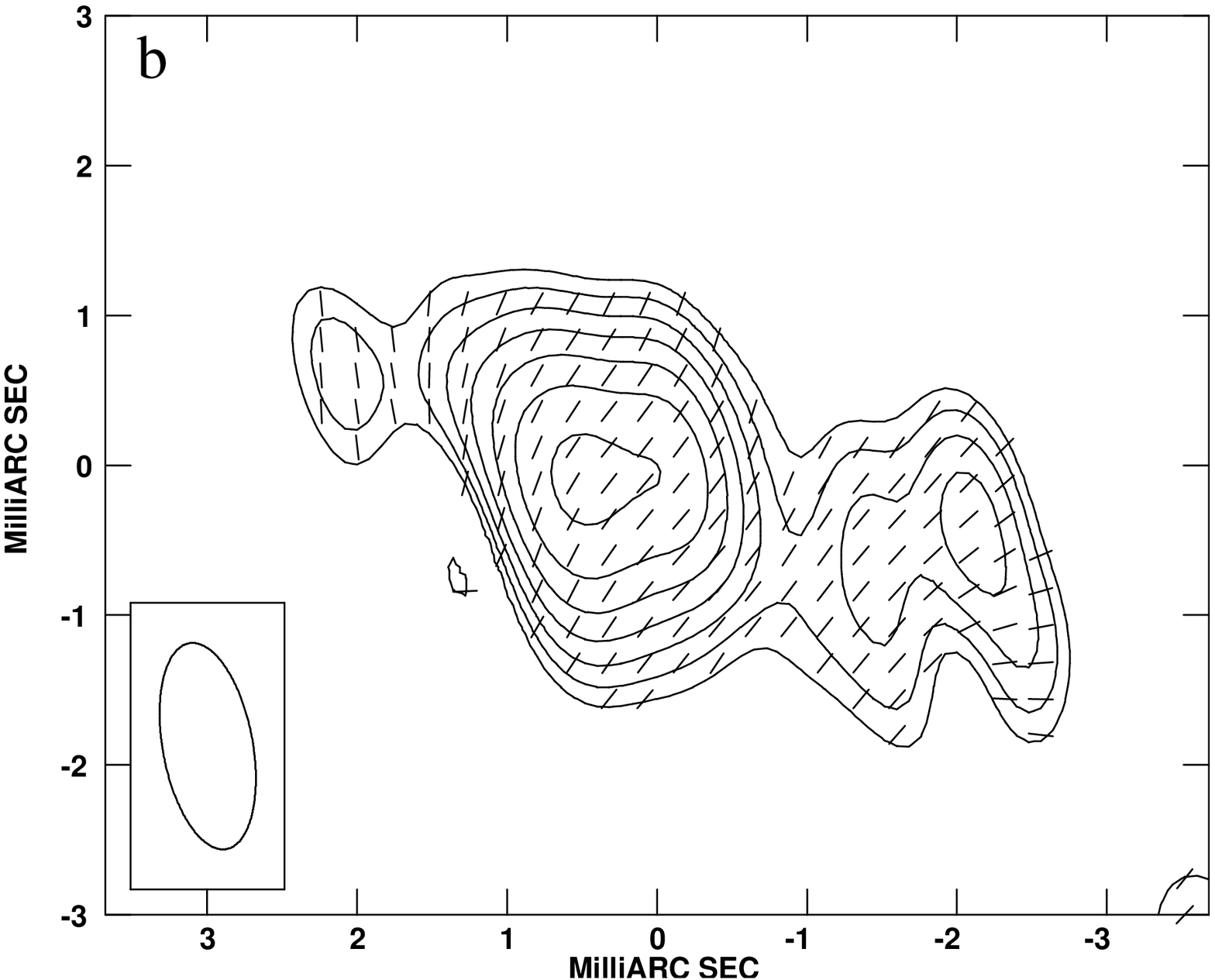,width=3.0in,angle=0} \\
\epsfig{file=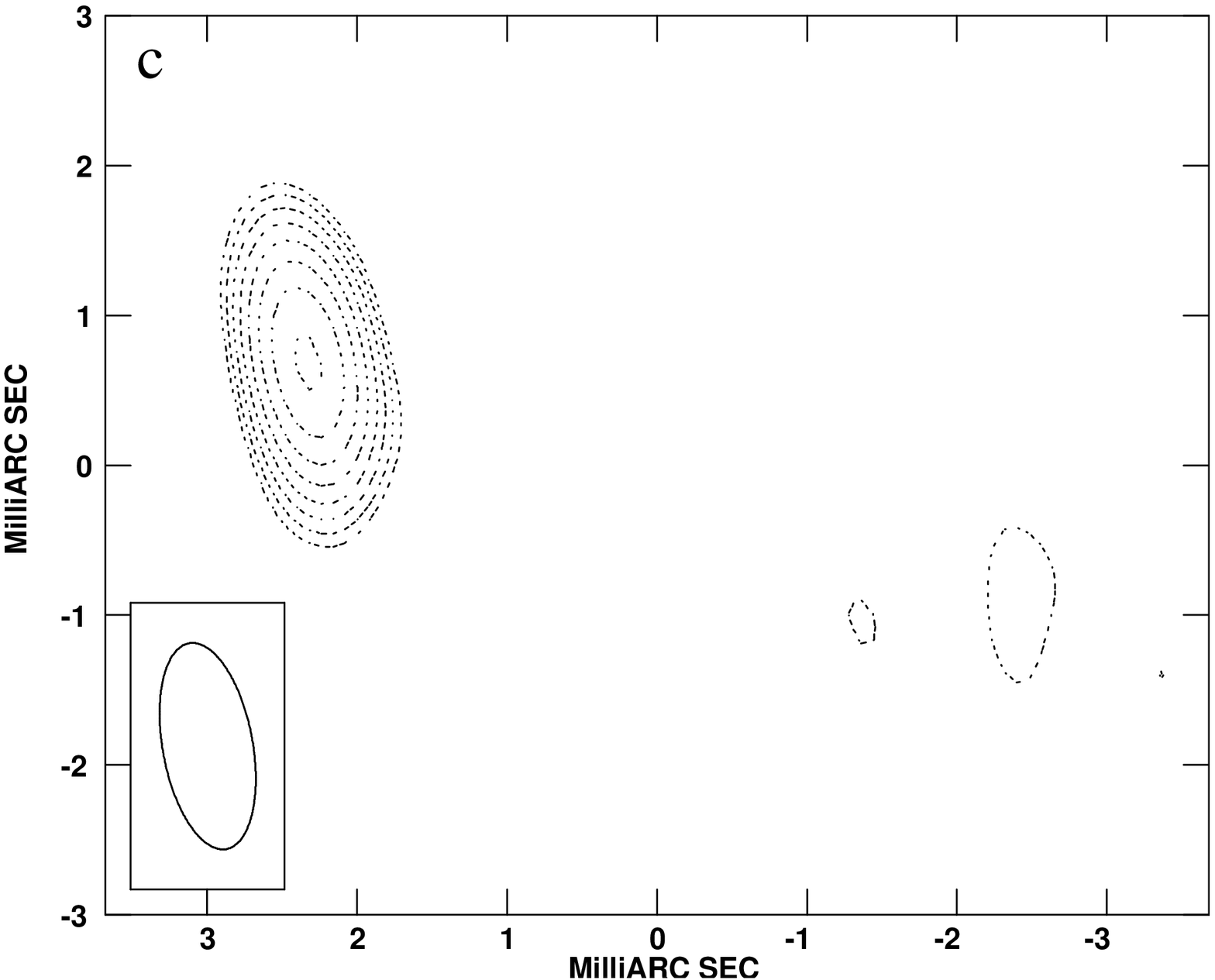,width=3.0in,angle=0}
\caption{\label{f:3c273c-ipv}
Naturally weighted images of 3C\,273, May 1996.  The images show
only the inner 5-6 mas of the jet
a) Total intensity $\sqrt{2}$ contours beginning at 0.030 Jy/beam.  The map
peak is 11.03 Jy/beam.  b) Linear polarization E-vectors superimposed on
polarization intensity with $\sqrt{2}$ contours beginning at 0.015 Jy/beam.
The map peak is 0.133 Jy/beam.  c) Circular polarization intensity
$\sqrt{2}$ contours beginning at $-$5 mJy/beam.  The map peak is $-$60
mJy/beam.}
\end{figure}

\placefigure{f:3c273c-ipv}

Figure~\ref{f:3c273c-ipv} displays the total intensity, linear 
polarization, and circular
polarization images of 3C\,273 for the 1996.41 epoch.  In these
images 3C\,273 has a strong core with a bright jet extending at a
position angle of $\sim -120^\circ$.  The core has weak linear
polarization at the $0.3\%$ level, and the jet is more strongly 
polarized at the $\simeq 10-40\%$ level.  The observed circular 
polarization lies directly on the core with an integrated
intensity of $-74$ mJy $\pm 15$ mJy ($m_c = -0.6\%$).

The total intensity
and linear polarization structure of the jet are very similar in the other
epochs.  The integrated flux density of the core, however, increases 
in strength from epoch 1996.05, where it is 2.8 Jy, to epoch 1996.57, where it
is 13.4 Jy.  By epoch 1996.74, the core is clearly
comprised of two components (denoted CE and CW in Table~\ref{t:3c273}).  We report
the results of u-v plane model-fitting of the core region for each epoch
in Table~\ref{t:3c273}.  

\placetable{t:3c273}

\begin{figure}
\epsfig{file=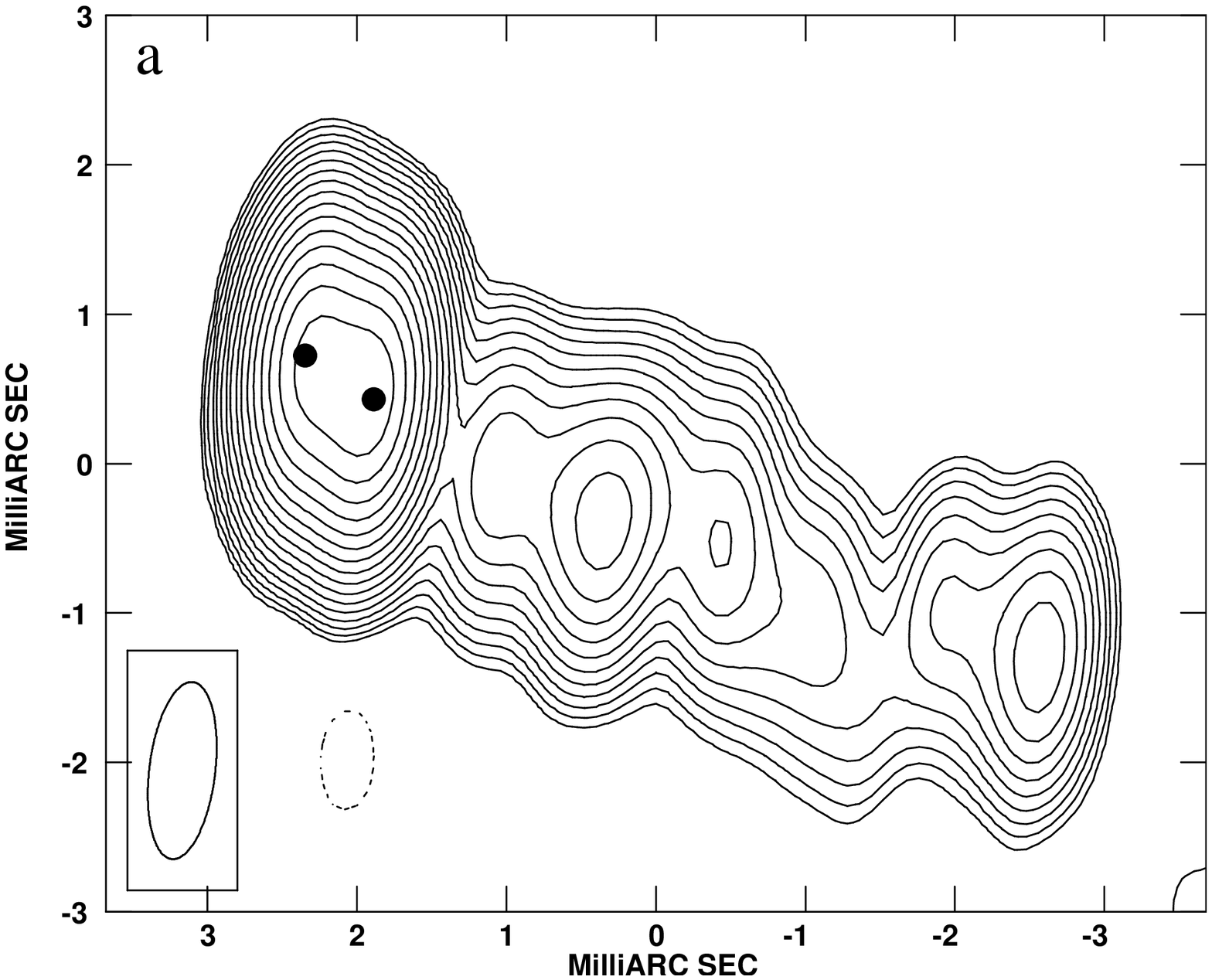,width=3.0in,angle=0} 
\epsfig{file=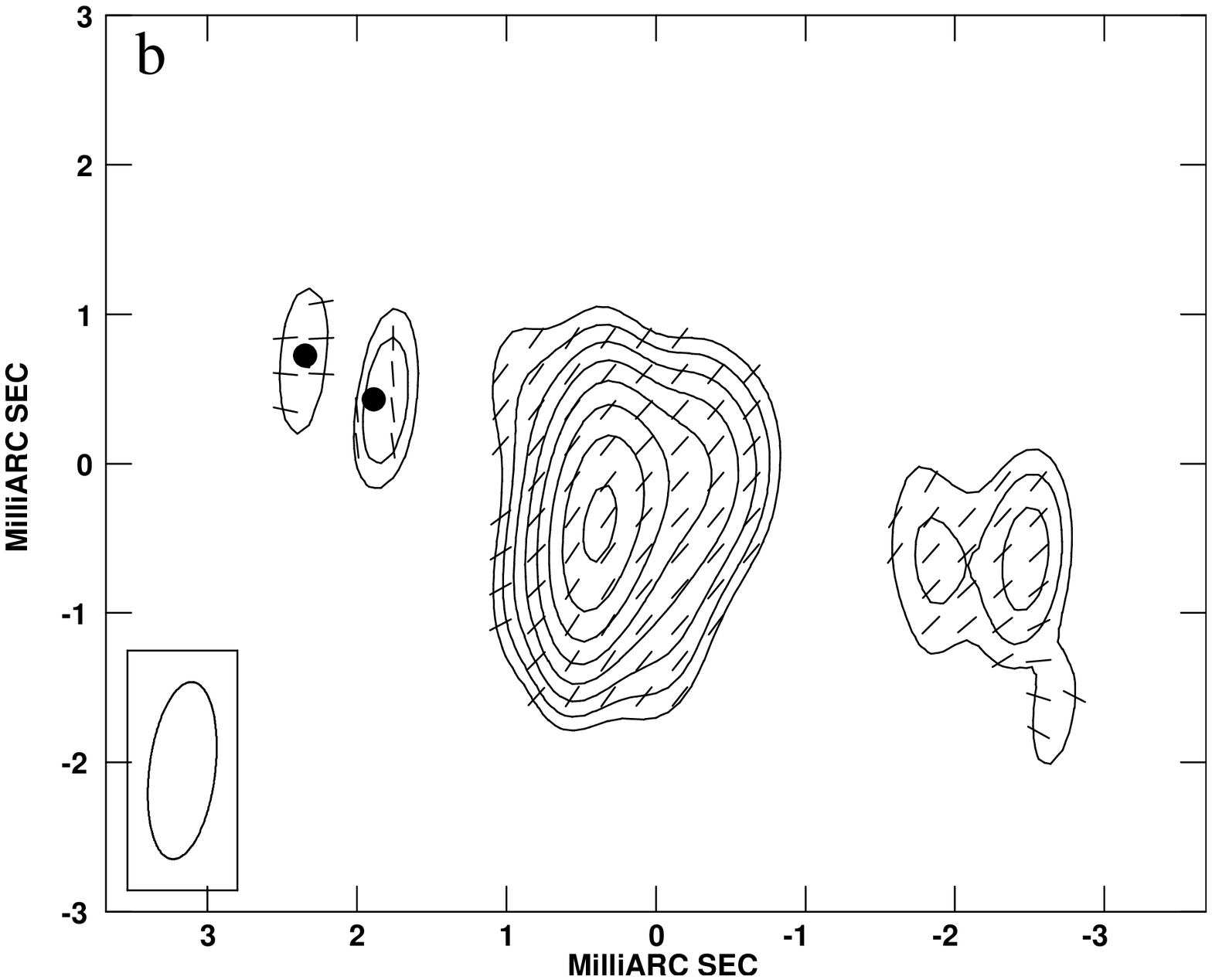,width=3.0in,angle=0} \\
\epsfig{file=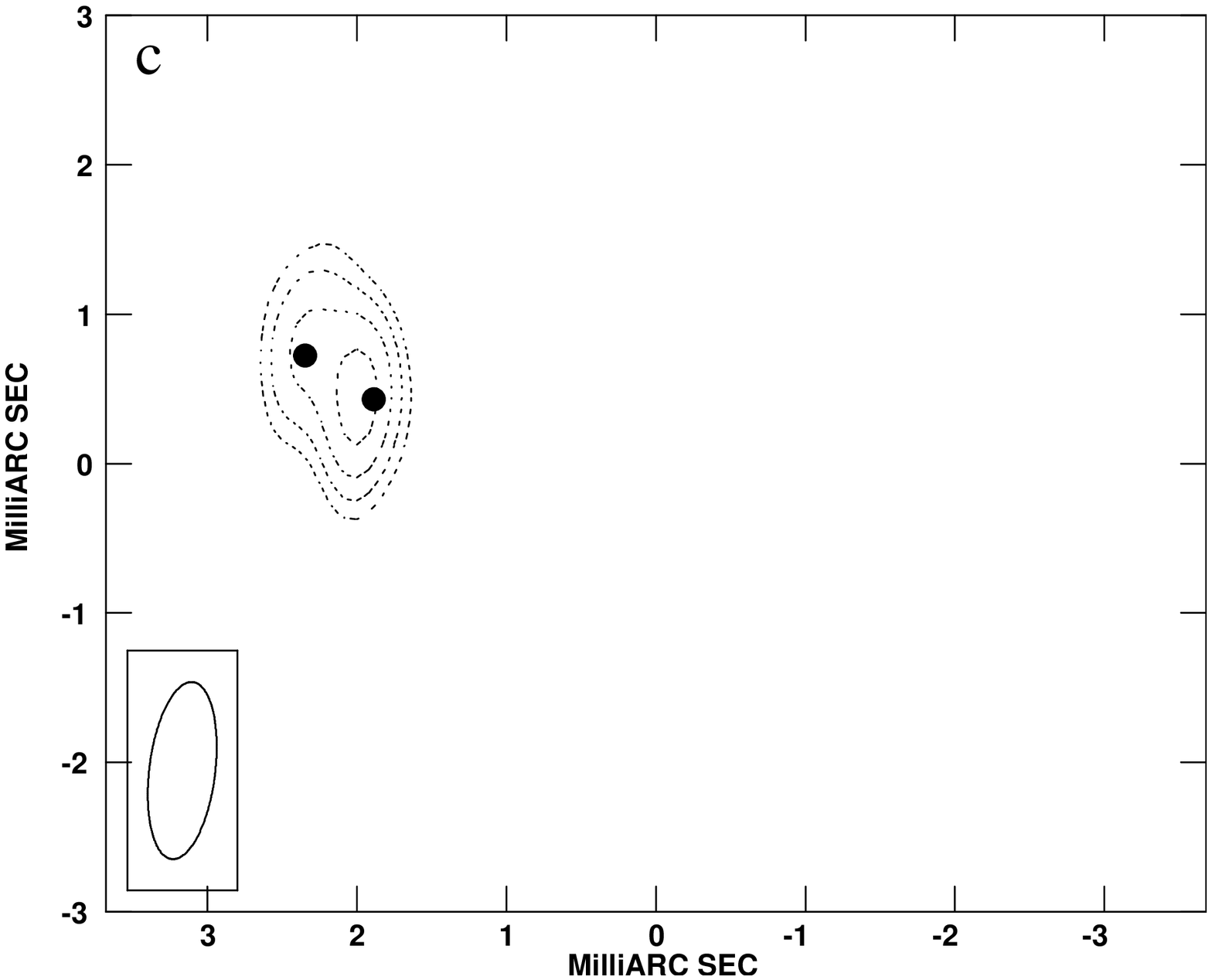,width=3.0in,angle=0}
\caption{\label{f:3c273e-ipv}
Uniformly weighted images of 3C\,273, September 1996.  The images show
only the inner 5-6 mas of the jet.  
a) Total intensity $\sqrt{2}$ contours beginning at 0.030 Jy/beam.  The map
peak is 7.59 Jy/beam.  b) Linear polarization E-vectors superimposed on
polarization intensity with $\sqrt{2}$ contours beginning at 0.015 Jy/beam.
The map peak is 0.184 Jy/beam.  c) Circular polarization intensity
$\sqrt{2}$ contours beginning at $-$10 mJy/beam.  The map peak is $-$34
mJy/beam.  Solid points on the images represent the locations of
the eastern and western components of the core. }   
\end{figure}

The circular polarization structure of the core also changes over the 
course of the epochs in a manner consistent with the total intensity 
and linear polarization changes.  In epoch 1996.05,  
the integrated circular polarization is $-10$ mJy $\pm 10$ mJy 
($m_c \sim -0.4$\% but consistent with zero) in
a very noisy {\em gain transfer} image.  At epoch 1996.23, the total
intensity of the core is much higher, but we detect no circular polarization 
with a limit $< 0.2$\%.  In 1996.41 (Figure~\ref{f:3c273c-ipv}), the
circular polarization is $-0.6$\% and the core also shows the first signs of
linear polarization.  In 1996.57, the circular polarization is $-0.5$\%
and the linear polarization has increased further.  By 1996.74 the core has split
in total intensity and linear polarization; the circular polarization also 
splits and is predominately associated with the western core (CW) component.
Figure~\ref{f:3c273e-ipv} displays uniformly weighted total intensity,
linear polarization, and circular polarization images from the 1996.74
epoch.

\placefigure{f:3c273e-ipv}

\begin{figure}
\begin{center}
\epsfig{file=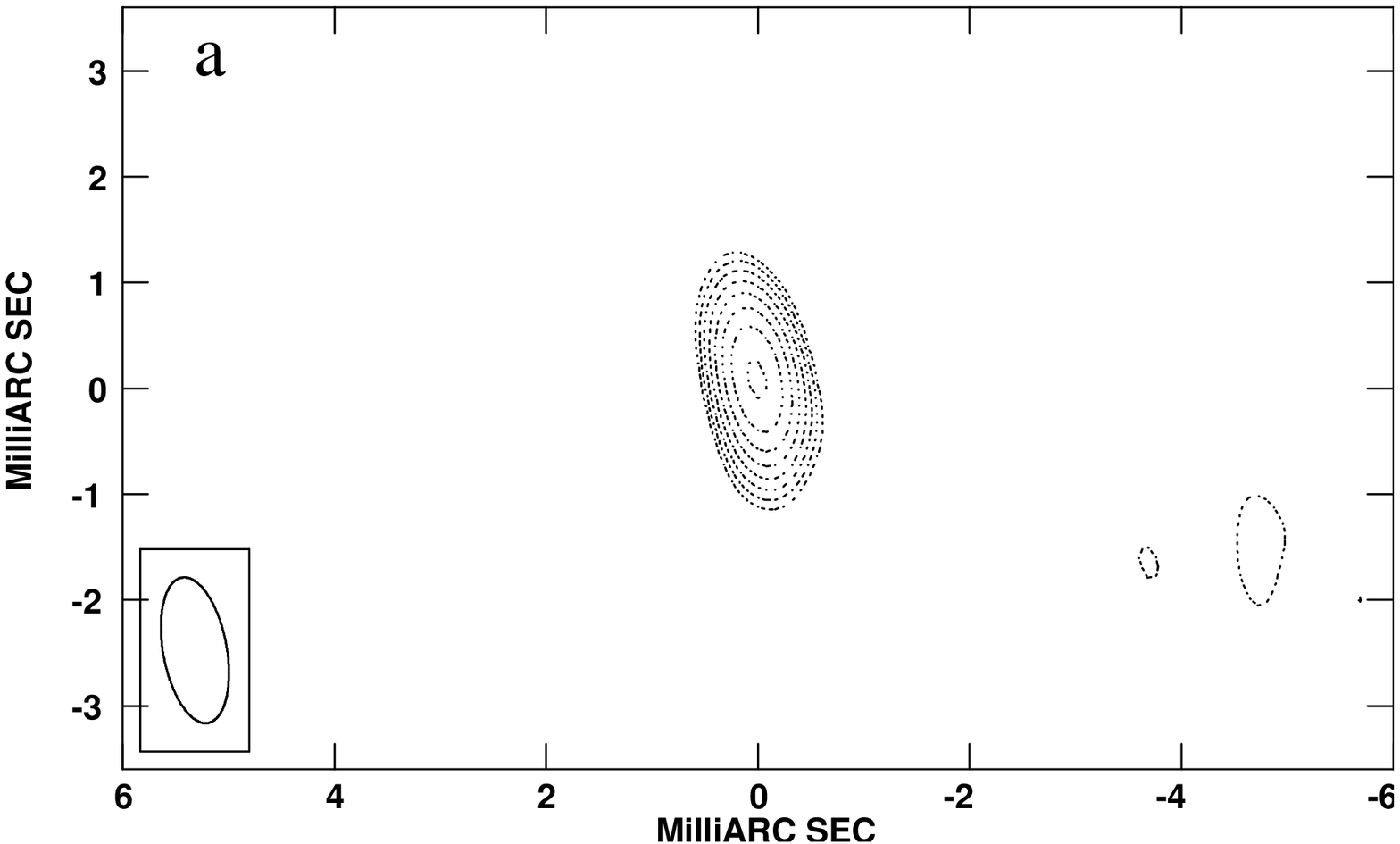,width=3.5in,angle=0} \\
\epsfig{file=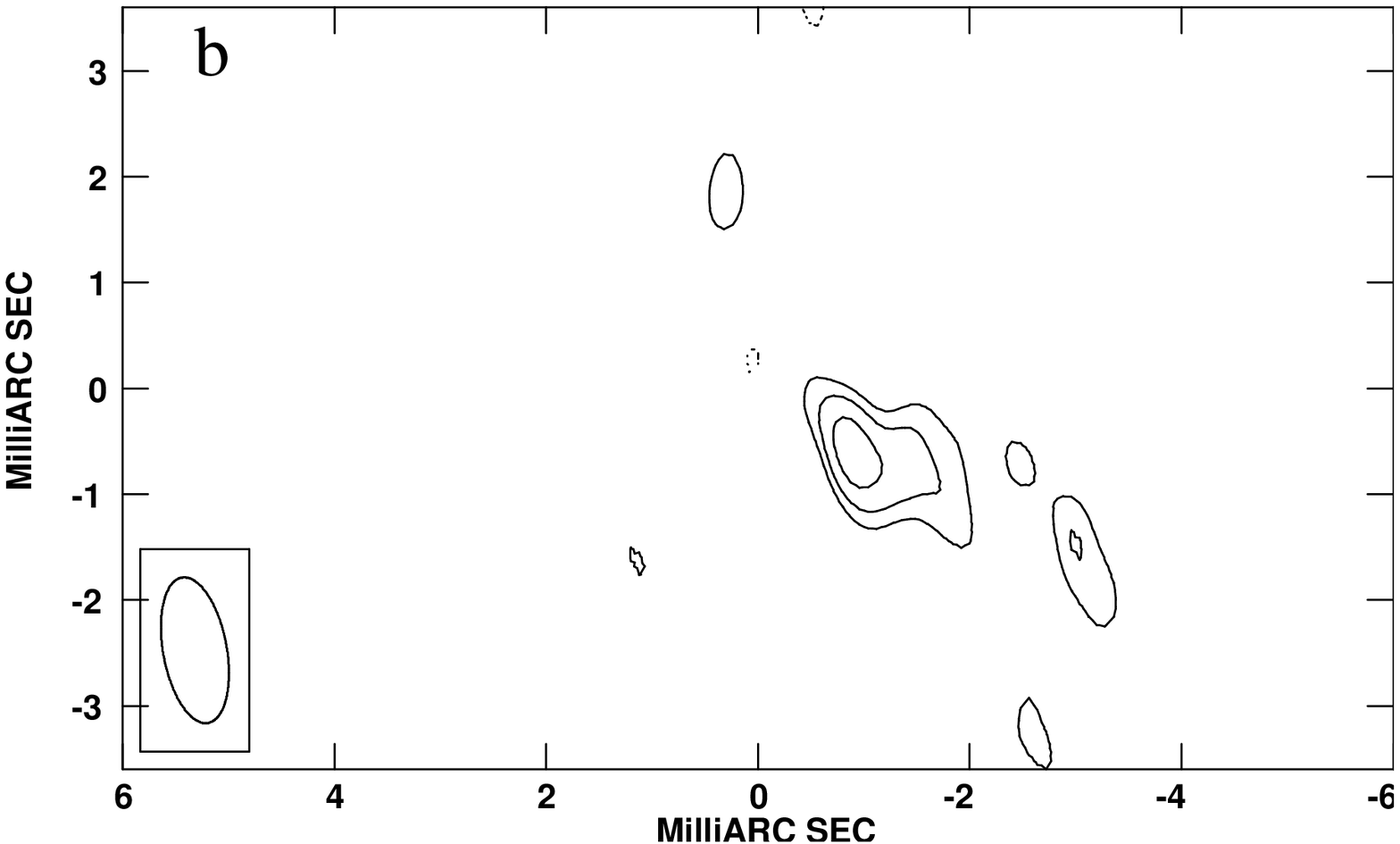,width=3.5in,angle=0} \\
\epsfig{file=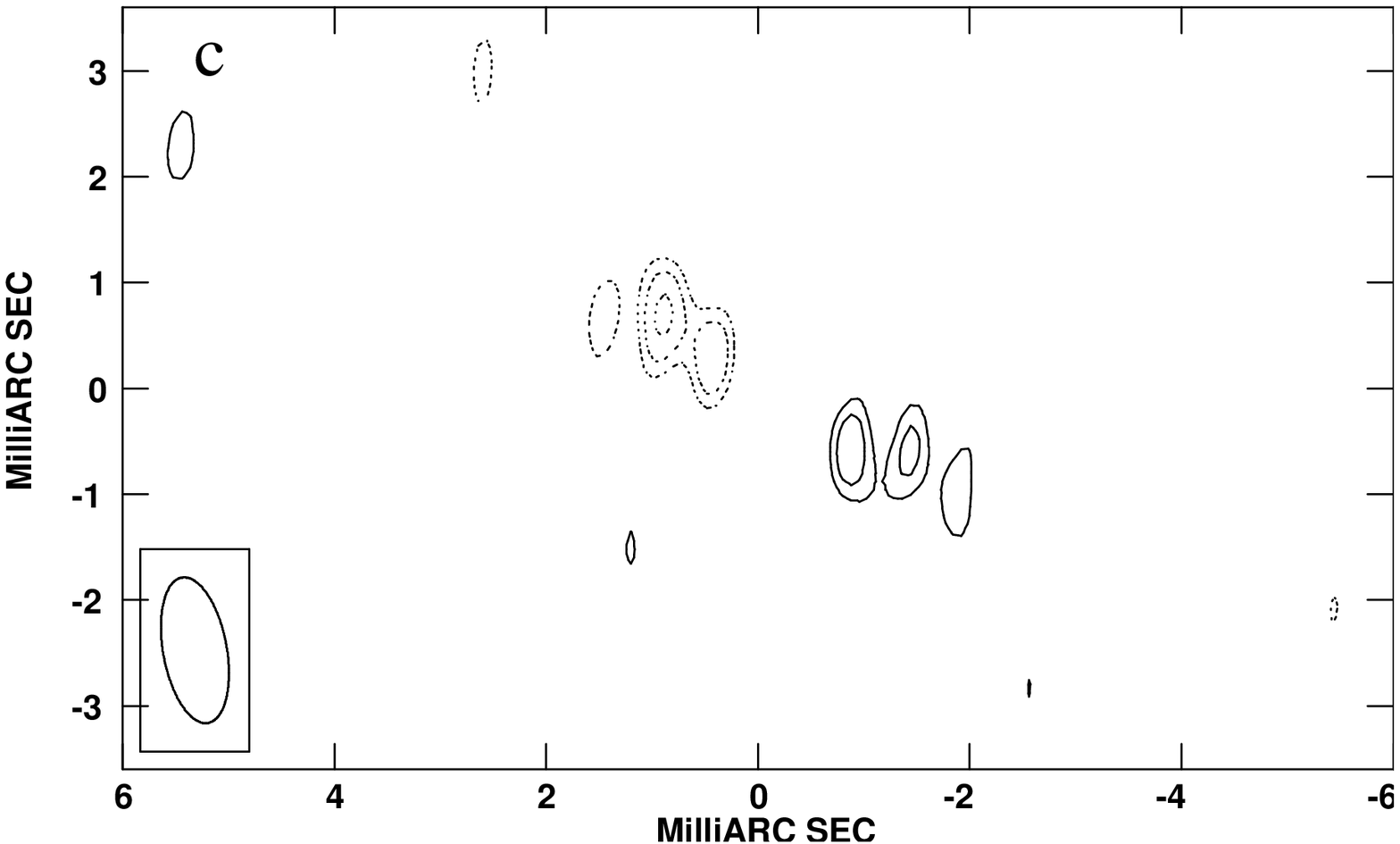,width=3.5in,angle=0}
\end{center}
\caption{\label{f:3c273c-3v}
Circular polarization images of 3C\,273, May 1996.  a) Gain
transfer calibration.  $\sqrt{2}$ contours begin at $-$5 mJy/beam.  The map
peak is $-$60 mJy/beam.  b) Zero-V self-calibration result.  $\sqrt{2}$
contours begin at +/$-$ 2.0 mJy/beam.  The map peak is +5.0 mJy/beam.  c)
Phase-only image.  $\sqrt{2}$ contours begin at +/$-$ 140 $\mu$Jy/beam.  The
map peak is +312 $\mu$Jy/beam.}
\end{figure}

\placefigure{f:3c273c-3v}

Figure~\ref{f:3c273c-3v} displays the circular polarization images produced by the {\em
gain transfer},  {\em zero-V self-cal}, and {\em phase-only} calibration
procedures for epoch 1996.41.  Figure~\ref{f:3c273c-3v}a is the 
{\em gain transfer} image presented in
Figure~\ref{f:3c273c-ipv}c, framed differently 
for ease of comparison to the other images.
Figure~\ref{f:3c273c-3v}b is the {\em zero-V self-cal} image.  It is clear that the
circular
polarization from the core has been transfered (by self-calibration) to the
jet with reversed sign and much reduced (absolute) level. 
Figure~\ref{f:3c273c-3v}c is the {\em phase-only} image.  As expected, the signal in the
phase-only map is anti-symmetric about the core with respect to the jet
location.  

Due to the complexity of 3C\,273, it is difficult to use the {\em zero-V
self-cal} and {\em phase-only} calibration techniques to derive firm
numbers for the true circular polarization of 3C\,273.  The best numbers
are derived from the {\em gain transfer} calibration procedure, but
it is clear that the {\em zero-V self-cal} and {\em phase-only} images
are consistent with the {\em gain transfer} results. 
The 1996.57 and 1996.74 
epochs also produce {\em zero-V self-cal} and {\em phase-only}
images consistent with the results from  
{\em gain transfer} calibration.

\subsection{Other Sources}

Circular polarization observations (using {\em gain transfer}
calibration) from our remaining nine sources are summarized in 
Table~\ref{t:rest}.  In most cases, these sources show no 
circular polarization from any of our calibration techniques.  The
upper limits and errors reported in Table~\ref{t:rest} are estimated 
to be roughly $\sqrt{2}$ times the noise peaks on
the circular polarization images.  
It is interesting
to note that, in general, the upper limits estimated from noise
peaks are consistent with the estimate of expected error of $\lesssim 0.15\%$
due to short term gain fluctuations (derived in \S{\ref{s:gain-trans}}). 

In isolated epochs, a few of the sources (J0738+17, J1512-09, J1751+09,
J1927+73, and J2005+77) appear to exhibit some core circular polarization.
In all of 
these sources, except J1927+73, the detected core circular polarization
is comparable to the peak noise in the image.  
In the three epochs in which it is detected, the circular
polarization signal for core of J1927+73 is significantly larger 
than the peak noise.  

\placetable{t:rest}
   
\section{Reliability of Calibration and Detection}
\label{s:discuss-rel}

In appendix~\ref{s:technical}, we analyze in detail the calibration
for circular polarization.  Here we provide a more general discussion
of the reliability of our calibration and claims of detection of circular
polarization.  

Chief among the sources of spurious circular polarization are antenna feed 
leakage (D-terms) and improper gain calibration.  Our analysis in \S{\ref{s:dterms}}
demonstrates that the D-terms, even if completely uncorrected, cannot be the 
source of the circular polarization we observe. Our correction of the D-terms is 
accurate enough to limit feed leakage errors in of circular polarization images
to $0.2$\% of the linear polarization (which itself is typically only a
few percent of the corresponding total intensity).  The antenna D-terms are not a 
factor in the reliability or quality of our results.

In \S{\ref{s:gain-cal}}, we describe the three gain calibration techniques 
we have used to detect and measure circular polarization.  Two of these techniques,
{\em gain transfer} and {\em zero-V self-cal}, derive independent 
measures of the R versus L antenna gains.  The third technique, {\em phase-only}
imaging, is independent of the antenna amplitude gains and is used to demonstrate
that a circular polarization signal is also present in the phases of the data.

The {\em zero-V self-cal} technique can reduce or relocate existing
circular polarization.  This technique assumes
that there is no true circular polarization on the source in question, i.e. that
$RR = \tilde{I}_{mod}$ and $LL = \tilde{I}_{mod}$.  
As a result, true circular polarization can be modeled 
in part or in whole as a gain error and removed from the data.  On a
point source any true circular polarization will be completely removed
by this technique; however, the same is not true for more complicated
sources.  Because true circular polarization is additive in the $RR$ and
$LL$ correlations,  it cannot be completely ``corrected'' by multiplicative
antenna gains.

For a typical source with a strong core and a much weaker jet, circular 
polarization from the core of the source will be {\em transfered} by 
the derived antennas gains (which try to minimize the overall circular 
polarization) to the weaker parts of the source, with reversed sign, and  at 
(roughly) the same local fractional level.
With the notable exception of 3C\,279, most sources are too complicated
to directly disentangle this effect and derive the true circular
polarization of the source. (Extended trial and error simulations
could, in principle, reconstruct the original circular polarization distribution.) 
Also, if the true circular polarization is located in the
weaker jet (as in 3C\,84), then {\em zero-V self-cal} does 
much less to modify it. 

The {\em gain transfer} technique provides our most direct measurement
of circular polarization.  The R/L gain ratios applied to 3C\,84, J0530+13,
3C\,273, and 3C\,279 were derived from the remaining nine sources.  
As a result, no assumption was made about the presence
of circular polarization on these four sources.  However, in this
procedure, the antenna gains applied to the other nine sources
are not really source independent.  To what extent can we trust the
results of the {\em gain transfer} on these sources?  

The sources in our observation were highly interleaved to maximize u-v 
coverage.  This fact, coupled with our four hour averaging interval
for the derived gain table, helps to make the gain table largely
independent of individual sources.  As described in 
 \S{\ref{s:gain-trans}}, we tested this assumption and found 
very nearly the same results when the sources which have circular 
polarization are allowed to contribute their gain corrections.
We conclude that the results on these other sources, which were
allowed to contribute to the gain table, can be trusted within the
quoted errors. 

\begin{figure}
\begin{center}
\epsfig{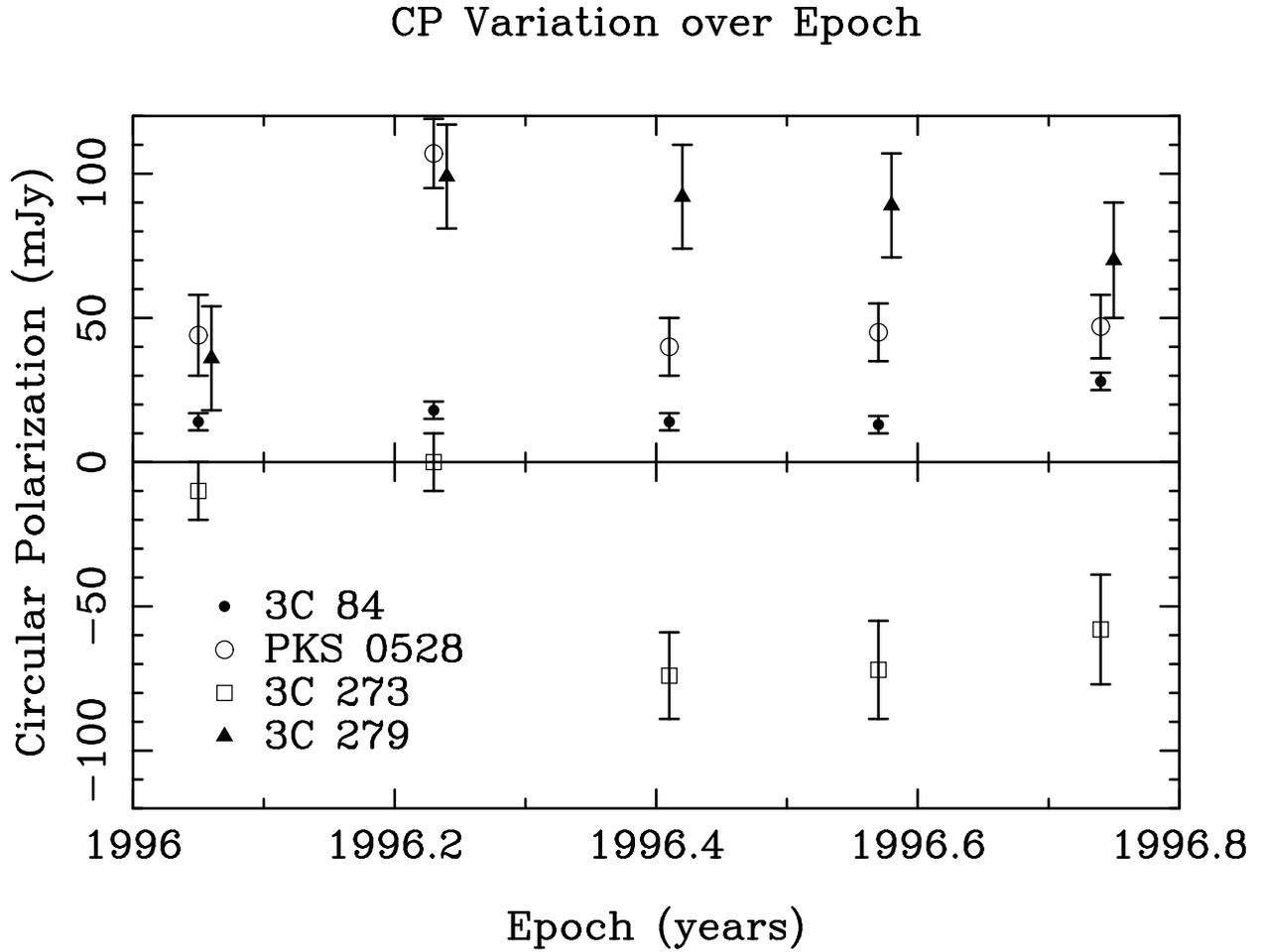}
\end{center}
\caption{\label{f:cp-v-epoch}
Circular polarization measurements plotted versus epoch.  All measurement
are from {\em gain transfer} calibration, excepting the 1996.57 epoch of
3C\,84 for which we plot the {\em zero-V self-cal} measurement.  The
data points for 3C\,279 are slightly displaced in time to avoid confusing
overlap with PKS 0528+134.}
\end{figure}

One danger in the {\em gain transfer} technique
is that the R/L antenna gains may not be stable on time scales long
enough to allow their transfer.  
Our results demonstrate that the gains are indeed stable enough to produce
reliable results.  The observed circular
polarizations are consistent in terms of sign, location, and amplitude from
epoch to epoch. (See Figure~\ref{f:cp-v-epoch}.)
We would not expect this consistency if the R/L antenna
gains were not stable on time scales of at least several hours.
It is important to note that the consistency is not perfect. This is 
partly due to source variability, but may
also reveal limits to the reliability of this technique.  In fact, our main
source of error in this technique is short time scale gain fluctuations which
go uncorrected.  Section~\ref{s:gain-trans} provides an estimate of
the errors from these fluctuations.

\placefigure{f:cp-v-epoch}

An important point is that the strong negative circular polarization
observed on 3C\,273 increases our confidence in the {\em gain transfer} 
calibration technique.  3C\,273 and 3C\,279 are neighbors in
the sky and interleaved in our observation schedule.  Due to the four hour 
averaging time, they will receive essentially the same set of $R/L$ antenna 
gains.  Any bias in the antenna gains would be expected to have the same 
effect on both sources; however, they have circular polarizations 
of {\em opposite} sign. 

The consistency between all of our techniques is best evaluated with 3C\,279
where we can use a simple point double model to directly compare the results
of all three approaches.  As demonstrated by our results in table \ref{t:3c279-2pt-ext}  
(graphically displayed in figure~\ref{f:3c279-2pt-ext}),  these techniques always
agree that a signal was detected and on the sign of that signal.  Furthermore, the
techniques agree in amplitude to within a factor of two (and often much better than
that). 

\section{Physical Interpretation}
\label{s:discuss-phy}

Circular polarization may be produced as an intrinsic component of 
synchrotron radiation or by Faraday conversion of linear to circular
polarization\markcite{JOD77} (e.g. Jones \& O'Dell 1977).  
Determining which mechanism
produces the circular polarization we observe is crucial to using
the observations as a physical probe of parsec-scale radio jets.  
Intrinsic circular polarization demands a significant uni-directional component of
magnetic field while Faraday conversion requires a significant 
population
of low energy relativistic particles in the jet\markcite{W98}
(Wardle et al. 1998).

In Wardle et al.~\markcite{W98}(1998) 
we analyze the 1996.57 epoch of 3C\, 279 in 
detail and use limits on the circular polarization present
in the 22 GHz companion observations to show that Faraday conversion is
the origin of the circular polarization observed in that source.
Our results set an upper limit on the low energy cutoff to the relativistic 
particle distribution of $\gamma_{min} \leq 20$ with likely values for 
$\gamma_{min}$ on the order of a few.  Using the results of Celotti
and Fabian\markcite{CF93} (1993), who showed that jets comprised 
mainly of an electron-proton
plasma must have $\gamma_{min} \geq 100$ to avoid carrying too much kinetic
energy, we suggested that 3C\,279 has a primarily electron-positron plasma. 

We cannot make similar arguments for the other sources until we have
their circular polarization results at 22 GHz.  However, we can make several 
observations about the presence of circular polarization in the sources 
we observe. 

In each of the sources where we have detected circular polarization, we 
find it either in or very near the core.  This is perhaps not too 
surprising given that the signals we detect are a very small fraction of
the total intensity and the core is the brightest part of the jet in
these sources. However,  in 3C\,84 we 
detect circular polarization not on the brightest part of the core, but rather
at the base of the southern jet, which is locally a factor of 2-3 less bright
than the map peak.  In 3C\,273 we do not detect significant circular polarization 
until epoch 1996.41, although the core is already very strong by 
epoch 1996.23.  By epoch 1996.74, the circular polarization of 3C\,273 is
predominately associated with the emerging core component, rather than
split evenly between both core components.  These results can be understood
from the work of Jones and O'Dell\markcite{JOD77} (1977), who find that circular 
polarization, however 
it is produced, is a strong function of opacity and is strongest near the $\tau = 1$
surface.

\begin{figure}
\begin{center}
\epsfig{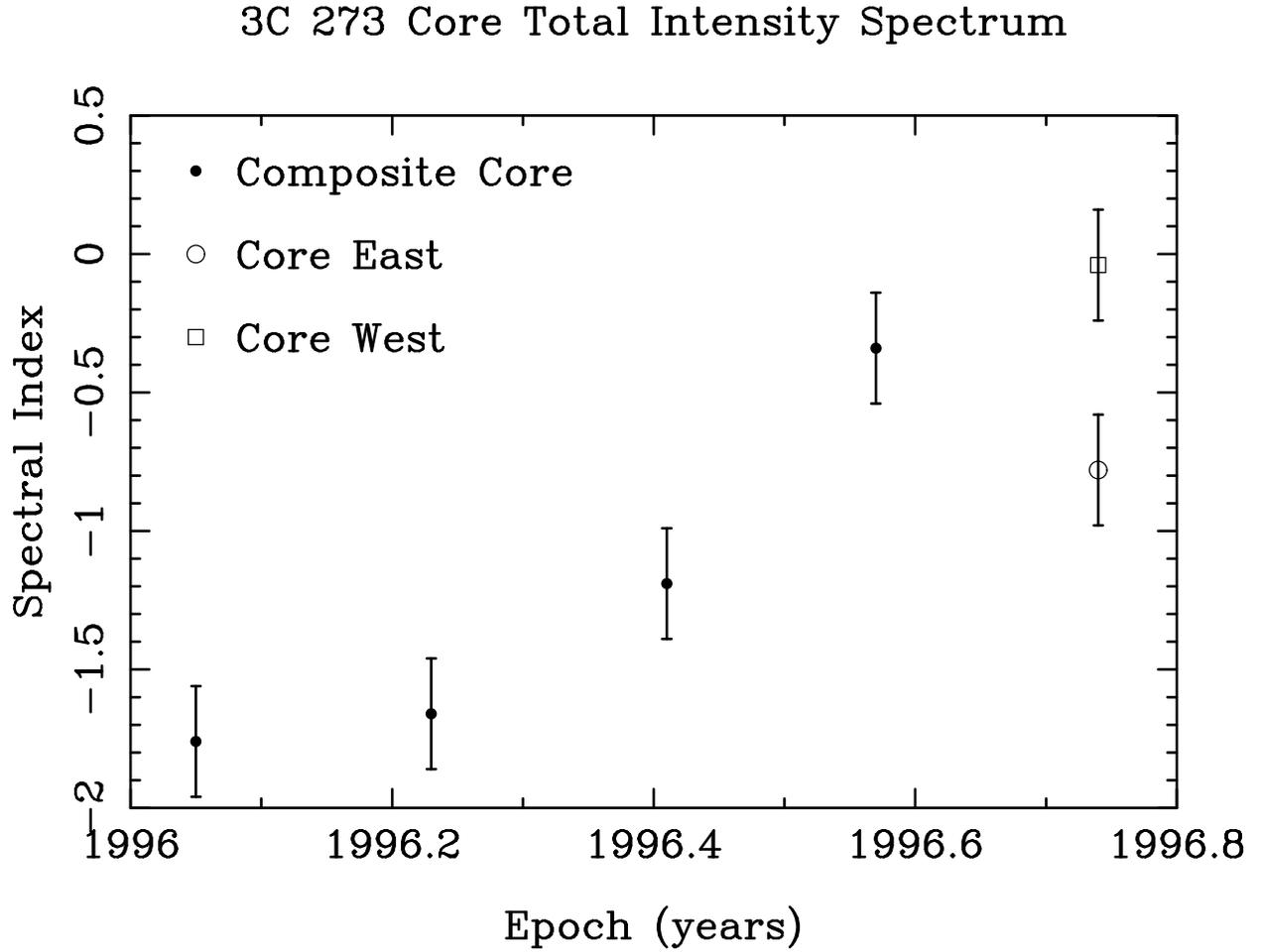}
\end{center}
\caption{\label{f:spec}
Spectral index, $\alpha$, of the core of 3C\,273 over time.  
$S \propto \nu^{-\alpha}$. $\alpha$ is measured between 15 and 22 GHz. 
In the fifth epoch, the core has separated into two pieces and the 
spectral index is plotted  for both.  ``Core West'' is the emerging 
component in that epoch.}
\end{figure}

The circular polarization we observe is linked to other properties of
the sources.  The core of 3C\,273 undergoes an outburst over the course
of our observations in 1996.  The May epoch observations 
of 3C\,273 are the first in which we see significant circular polarization in 
this source.  May is also the first epoch in which we observe linear 
polarization in the core of 3C\,273.  The simultaneous appearance of linear
and circular polarization in the core of 3C\,273 most likely results from
the changing opacity in the core region.  Figure~\ref{f:spec} shows the 
spectral index of the core of 3C\,273 between 15 and 22 GHz 
as a function of epoch\markcite{O98} (Ojha 1998).  This figure demonstrates that
the core begins to become significantly less opaque by the 1996.41 epoch
when we first detect strong circular polarization.

\placefigure{f:spec}

The core of 3C\,279 also undergoes an outburst in 1996.  
We observe the western core component, CW, to nearly triple in strength 
from January to September.  The circularly polarized flux we observe in 
3C\,279 appears to track the rise in total flux of this component, 
although the uncertainty
of the measurements in the 1996.23 and 1996.41 epochs makes detailed
analysis of the evolution difficult.

A second result is that the {\em sign} of the 
circular polarization is constant across epoch in all of our
sources. 3C\,84, PKS 0528+134, and 3C\,279 all have positive
circular polarization while 3C\,273 has negative circular
polarization.  3C\,84 changes very little in total intensity
or linear polarization over our observations, but the other 
three sources undergo significant changes in flux and/or linear
polarization.  For the sign of the circular polarization to remain
constant, there must be a stable, uni-directional 
component of magnetic field in the jet.  

In fact, both Faraday conversion and intrinsic circular
polarization require the presence of some excess uni-directional 
component of magnetic field.  In the case of conversion, this field 
provides the small amount of Faraday rotation necessary for
conversion. (Conversion cannot operate on pure
Stokes Q.)  A tangled field geometry or differential abberation 
\markcite{JPC} (Jones, personal communication) can remove the
need for Faraday rotation, but such a geometry is unlikely to
produce the sign consistency we observe. 
 It is also possible to produce some net uni-directional field 
stochastically, but again, it
would not be expected to remain constant (in sign) while a source
is undergoing large internal changes.    

Single dish circular polarization observations at 8 GHz by Hodge and 
Aller\markcite{HA77} (1977) detected about $+0.1$\% 
integrated circular polarization on 
3C\,279, consistent in sign with our observations.  They detected 
about $-0.1$\% integrated circular polarization on 3C\,273, also consistent 
with our observations.  For 3C\,84, they measured about $-0.1$\% circular
polarization which is opposite in sign to our observations.  Future multi-frequency
VLBI observations of 3C\,84 will be important to determine if this sign 
difference is an opacity effect or reflects slow changes in the source. Their observations
of 3C\,279 and 3C\,273 suggest that the uni-directional component of magnetic field in
these sources may be stable on a time scale of decades in our frame of
observation.  It is possible that this is directly related to the magnetic field at the
central engine\markcite{BBR84} (e.g. Begelman, Blandford, \& Rees 1984). 

It is also interesting to examine
the relation of circular polarization to linear polarization.  
In the cores of 3C\,84 and 3C\,273
we observe equal or {\em less} linear polarization than circular polarization.
The linear polarization observed, if taken as a direct measure of field order, is
not enough to produce the observed circular polarization by Faraday conversion.  
Lack of observed linear polarization also presents a problem if the 
circular polarization is produced by the intrinsic mechanism which requires a 
significant uniform field.  The most likely explanation is that external Faraday 
depolarization in the nuclear region of the AGN has significantly reduced the 
degree of linear polarization without
affecting the circular polarization.  This seems 
to be the case for 3C\,84\markcite{W71} (Wardle 1971), and we note 
that Taylor\markcite{T98} (1998)
reports a large rotation measure of -1900 rad/m$^2$ for the core of 3C\,273.

\section{Conclusions}
\label{s:conclusions}

We have observed parsec-scale circular polarization in the radio
jets of 3C\,84, PKS 0528+134, 3C\,273, and 3C\,279 with the VLBA at 15
GHz.  
For each source our detections are at the level of $\sim 0.3-1\%$ local
fractional circular polarization.  Each individual detection has
a significance in the range of 3 to 10 $\sigma$. The circular
polarizations we observe are consistent across epoch with changes in 
structure clearly linked to physical changes within the sources.
To confirm our detections, we have devised three calibration techniques to
help mitigate our sensitivity to gain errors.  We find
that all three techniques give consistent and reproducible results.

Our observations of circular polarization are closely tied to 
the physical conditions within the radio jets.  We always detect
circular polarization near the region of unit optical depth.  In
two sources, 3C\,273 and 3C\,279, we observe core outbursts during
1996 and find that the circular polarization is closely tied to 
these events.  In 3C\,273 we first observe significant circular polarization 
in the core as the outburst is reaching its maximum, simultaneous with
our first observations of linear polarization from the same region.  When
a new component emerges from the core of 3C\,273, the circular
polarization is predominately associated with it.  

Circular polarization may be produced as an intrinsic component 
of synchrotron radiation or by the Faraday conversion of linear to circular 
polarization.  Our observations, at a single frequency, do not distinguish 
between these possible mechanisms, 
but independent of the mechanism, the remarkable consistency across epoch of 
the sign of the observed circular polarization suggests the existence of a 
long term, stable, uni-directional component of the magnetic field.  Single dish 
observations of 3C\,273 and 3C\,279 at 8 GHz by Hodge and Aller (1977) suggest that 
this stability in these sources may span decades in our frame of observation.  

Direct imaging of circular polarization with the VLBA provides an entirely
new probe of the magnetic field structure and particle spectrum of the
parsec-scale radio jets of AGN.  Multi-frequency observations of circular 
polarization will be able to distinguish between the intrinsic and
Faraday conversion mechanisms for producing circular polarization.  By
imaging all four Stokes parameters at multiple frequencies, we can construct 
detailed models of the entire radiative emission and transfer through
the source and begin to determine the composition and energy spectrum of the 
relativistic plasma within the jet.  

\section{Acknowledgments}

We would like to thank R. Ojha for a careful reading of the manuscript.  We
also thank Tom Jones and Craig Walker for helpful discussions. 
This work has been supported by NSF Grants AST 92-24848, AST 95-29228, 
and AST 98-02708 and NASA Grants NGT-51658 and NGT5-50136.


\appendix

\section{Calibration for Circular Polarization}
\label{s:technical}

The VLBA is equipped with circularly polarized
feeds, right ($R$) and left ($L$) at each antenna.  In the absence of 
instrumental effects, the parallel hand correlations are linear combinations
of the total intensity ($I$) and circular polarization ($V$).
\begin{equation}  
R_iR_j = \tilde{I}_{ij} + \tilde{V}_{ij}
\end{equation}

\begin{equation}  
L_iL_j = \tilde{I}_{ij} - \tilde{V}_{ij}
\end{equation}

As $\tilde{V}$ is typically $\leq 0.01\tilde{I}$, 
to detect circular polarization we must measure the small difference
between two large quantities and are therefore very sensitive to
calibration errors.  It is critical to understand and
remove effects which may corrupt a circularly polarized signal or
may produce spurious circular polarization.  These effects include
instrumental polarization, gain errors, beam squint, and baseline
based effects.

\subsection{Instrumental Polarization}
\label{s:dterms}

In the leakage feed model\markcite{RWB94} (e.g. Roberts et al. 1994)
the output antenna voltages for nominally right and left circularly 
polarized feeds are given by
\begin{equation}
V_{R} = G_{R}(E_{R}e^{-i\phi} + D_{R}E_{L}e^{+i\phi})
\end{equation}
\noindent and
\begin{equation}
V_{L} = G_{L}(E_{L}e^{+i\phi} + D_{L}E_{R}e^{-i\phi})
\end{equation}

\noindent where $G_{R}$ and $G_{L}$ are complex gains, $D_{R}$ and $D_{L}$ are 
the complex, fractional response of each feed to the orthogonal polarization, and 
$\phi$ is the parallactic angle.

The time-averaged, parallel-hand complex correlations of these voltages 
detected at two antennas can be expressed\markcite{RWB94} (Roberts et al. 1994) 
in terms of the stokes parameters as:
\begin{equation}
R_iR_j^* = G_{iR}G_{jR}^*[(\tilde{I}_{ij} +
\tilde{V}_{ij})e^{i(-\phi_i+\phi_j)} + 
D_{iR}D_{jR}^*(\tilde{I}_{ij} - \tilde{V}_{ij})e^{i(\phi_i - \phi_j)} 
\end{equation}
\[+ D_{iR}\tilde{P}_{ji}^*e^{i(\phi_i+\phi_j)} + 
D_{jR}^*\tilde{P}_{ij}e^{-i(\phi_i+\phi_j)}] \]
\noindent and
\begin{equation}
L_iL_j^* = G_{iL}G_{jL}^*[(\tilde{I}_{ij} - \tilde{V}_{ij})e^{i(\phi_i-\phi_j)} + 
D_{iL}D_{jL}^*(\tilde{I}_{ij} + \tilde{V}_{ij})e^{i(-\phi_i + \phi_j)}
\end{equation}
\[+ D_{iL}\tilde{P}_{ij}e^{-i(\phi_i+\phi_j)} + 
D_{jL}^*\tilde{P}_{ji}^*e^{i(\phi_i+\phi_j)}] \]
\noindent where $\tilde{P}$ is the linear polarization.

To detect circular polarization it is necessary to correct the
parallel-hand correlations for the D-term leakage.  This is particularly 
important for sources with high degrees of linear polarization.  Suppose the 
measured D-terms are related to the 
true D-terms in the following way.
\begin{equation}
D^{true} = D^{measured} + \Delta
\end{equation}

The leakage of D-term errors 
into the circular polarization image can be approximated by 
\begin{equation}
V_{leak} \sim  
\frac{I}{\sqrt{N_sN_a(N_a-1)}}(2\Delta D^{meas} +\sqrt{2(N_a-1)}m_L\Delta  + \Delta^2) 
\end{equation}
\noindent where $\Delta$ and $D^{meas}$ are rms values, $m_L$ is the fractional linear 
polarization, $N_a$ is the
number of antennas, and $N_s$ is the number of scans which are separated in
parallactic angle. A factor for the number of IFs is omitted because the D-terms
are strongly correlated between IFs.  In the worse case scenario of no
D-term correction
\begin{equation}
V_{leak} \sim \frac{I}{\sqrt{N_sN_a(N_a-1)}}(\sqrt{2(N_a-1)}m_LD + {D^2}).
\end{equation}

\begin{figure}
\begin{center}
\epsfig{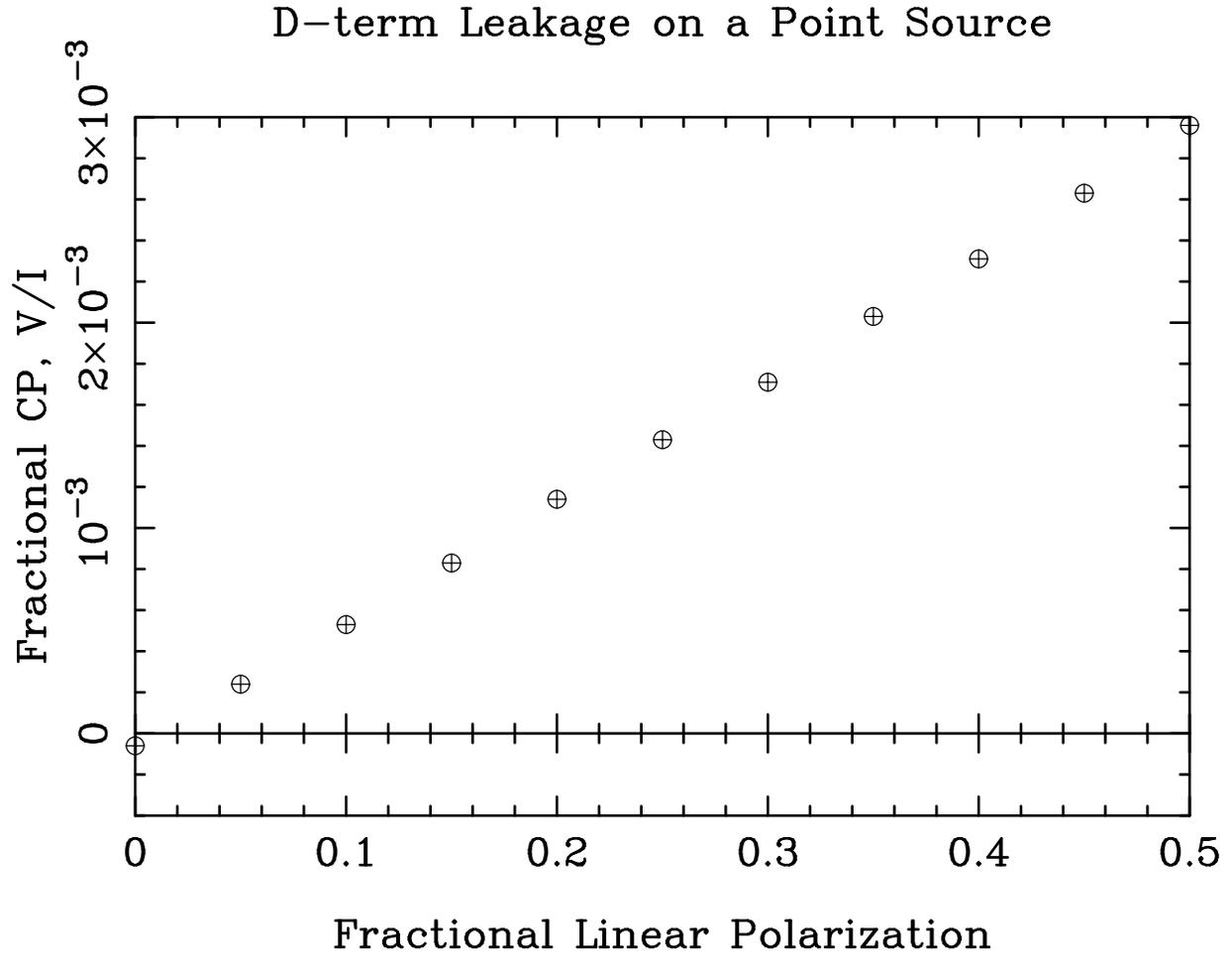}
\end{center}
\caption{\label{f:leak}
$\frac{V}{I}$ vs. fractional linear polarization 
for a point source 
with no intrinsic circular polarization.  Spurious circular polarization
results from leakage of uncorrected
2.5$\%$ D-terms.}
\end{figure}

Figure~\ref{f:leak} is a plot of $\frac{V}{I}$ versus fractional polarization, 
$m_L$, for 
a point source with no intrinsic circular polarization.  The D-terms have
an RMS of $2.5\%$ and were added to an originally D-term free point model 
created by the AIPS task UVMOD.  The model was built on a 5 scan, 10 antenna
observation.  The D-terms were added by `applying' the D-term
solutions found for a recent 15 GHz VLBA observation.  
The ratio, $\frac{V}{I}$, was measured at the location of the I peak in the cleaned
images. The slope and intercept of this plot agree well with the
expected relationship.

\placefigure{f:leak}

For the observations presented in section~\ref{s:images}, the measured D-terms
are typically $1$-$5$\% with $\sim2$\% being most common.  We estimate their errors 
to be about $15$\% of the measured
D-terms, so $\Delta \sim 0.2$ to $0.8$\%.  These errors were estimated by
comparing the D-term solutions from PKS 0528+134 (which were applied to all
sources) to D-term solutions found independently on other sources in the
same experiment.  With the D-terms corrected to this level of accuracy, 
$V_{leak}/I < 0.002m_L + 3\times 10^{-5}$ in our observations. 

\subsection{Gain Calibration}
\label{s:gain-cal}
Correct gain calibration of the R and L complex gains at each antenna is
critical to the detection and measurement of circular polarization.  In
particular, it is the calibration of the gain difference between the R
and L feeds, best represented by the complex gain ratio, R/L, which 
presents the largest problem for circular polarization observations using
circular antenna feeds.

Self-calibration attempts to correct the complex antenna gains by comparing
the u-v data to the transform of an idealized model of the source.  
These solutions are found in a least-square, antenna-based manner with all
of the baselines to a particular antenna used to constrain the solution.
Ideally, the solutions for all antennas are found simultaneously.

For simplicity, assume that we have the correct total intensity model for
a source, $I_{mod}$. The corresponding data may then be self-calibrated using
$I_{mod}$ in one of two ways:

\begin{itemize}
\item Assuming $RR = \tilde{I}_{mod}$ and $LL = \tilde{I}_{mod}$.  This method 
      derives a separate correction for the Right and Left hand systems 
      at each antenna.
\item Assuming $(\frac{RR + LL}{2}) = \tilde{I}_{mod}$ and thus deriving a
      single correction to be applied to both the Right and Left hand
      systems at each antenna.
\end{itemize}

\subsubsection{Self-Calibration Assuming Zero Circular Polarization.}
\label{s:zerov}

The first method derives a separate correction for the right and left hand 
systems at each antenna but assumes no circular polarization in the source.  
In principle this method can correct the R/L
complex antenna gain ratios in the absence of real circular polarization.  In
the presence of real circular polarization the complex correlations are
\begin{equation}
R_iR_j^* = G_{iR}G_{jR}^*(\tilde{I}_{ij} + \tilde{V}_{ij})
\end{equation}
\noindent and
\begin{equation}
L_iL_j^* = G_{iL}G_{jL}^*(\tilde{I}_{ij} - \tilde{V}_{ij}).
\end{equation}

The measured gains that the second technique finds and removes from the
data will be related to the true complex gains by
\begin{equation}
G_{iR}^{true} = G_{iR}^{meas}g_{iR}
\end{equation}
\noindent where the $g$'s are residual gains that satisfy the conditions
\begin{equation}
g_{iR}g_{jR}^*(\tilde{I}_{ij} + \tilde{V}_{ij}) \approx \tilde{I}_{ij}
\end{equation}
\noindent and
\begin{equation}
g_{iL}g_{jL}^*(\tilde{I}_{ij} - \tilde{V}_{ij}) \approx \tilde{I}_{ij}
\end{equation}
\noindent as well as possible across the array.  Real circular polarization
is additive in the $RR$ and $LL$ correlations, so the multiplicative gains
applied by self-calibration cannot completely remove real circular polarization
except when the source is a simple point in total intensity.  In fact, 
this kind of gain calibration may reduce and relocate real circular
polarization to weaker components but {\em cannot create it}. 

\begin{figure}
\begin{center}
\epsfig{file=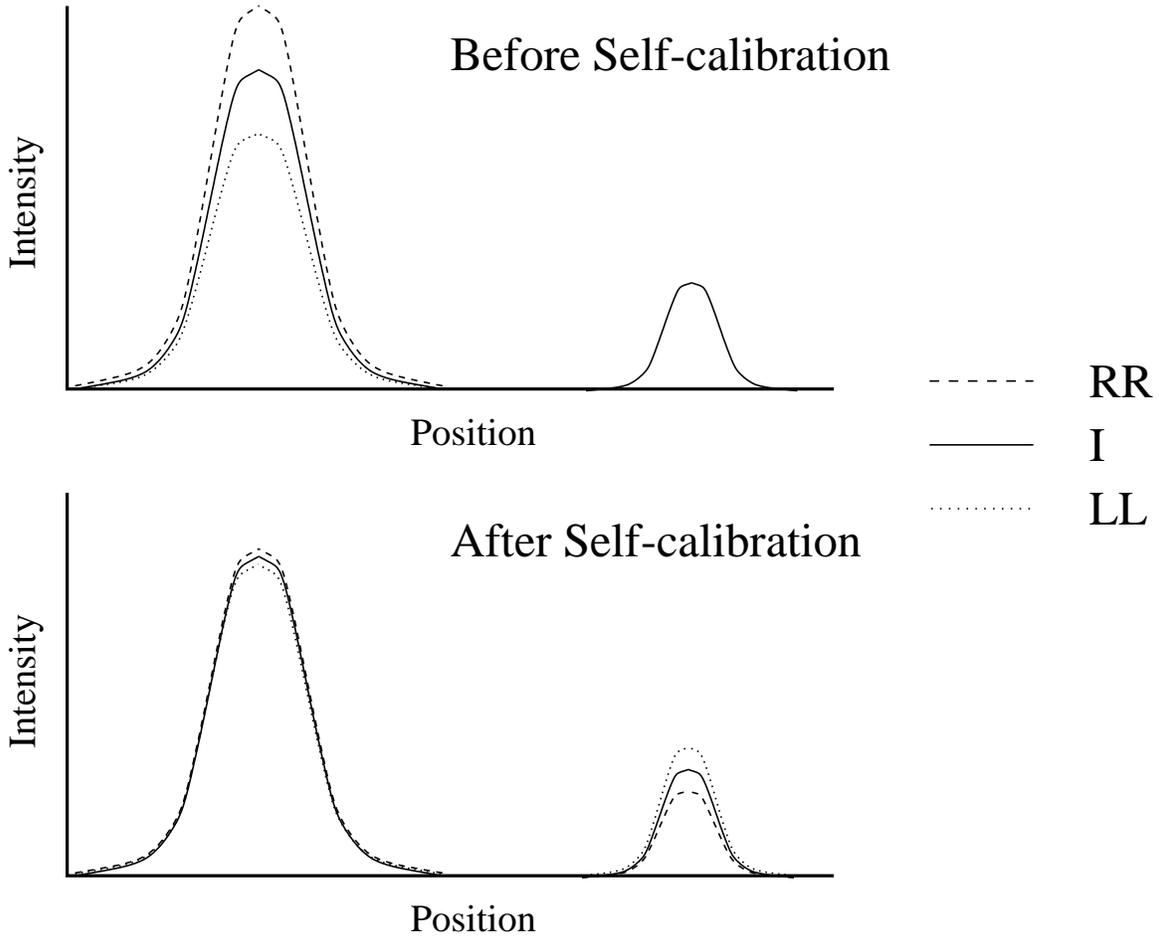,width=6.0in,angle=0}
\end{center}
\caption{\label{f:zerov}
Simple graphical model of the effect of self-calibration
assuming no circular polarization.  The amplitudes
of $RR$, $LL$, and $I$ before and after self-calibration are illustrated.  
$V = (RR-LL)/2$.  
Before self-calibration, there is significant circular polarization
on the strong $I$ component.  Self-calibration assuming no $V$ will force $RR$
and $LL$ to agree with $I$ as closely as possible by adjusting the antenna gains.
Circular polarization, however, is not multiplicative but additive in the $RR$ 
and $LL$ correlations.  Adjusting the antenna gains to remove the circular polarization 
on the strong component will induce circular polarization with the opposite sign on the 
weak component.  So, after self-calibration, the weak component has negative 
circular polarization at roughly the same fractional level as the original 
positive circular polarization on the strong component.}
\end{figure}

\placefigure{f:zerov}

Self-calibration with the assumption of zero circular polarization  
can seriously modify any circularly polarized signal present in a data set.
Circular polarization originating on weak components seems 
to be largely unmodified,
while circular polarization originating on strong components 
is shifted (with reverse sign but the same fractional level) 
to the weaker components in the $I$ image. Figure \ref{f:zerov} shows a
point double source with real circular polarization, before and after
self-calibration.

For point sources, such as PKS 0528+134, real circular polarization will look
simply like a difference in amplitude gain between $RR$ and $LL$ and will be 
completely removed by this procedure.

For an unequal point double, such as 3C\,279, there is an inherent ambiguity 
about the true location of any circular polarization detected with 
{\em zero-V self-cal}.  Circular polarization may have originated on either the weak
or the strong component but will show up only at the location of the weak component
in the {\em zero-V self-cal} image.  Assuming the circular polarization 
originated on the stronger core component, it is possible to 
extrapolate the true circular polarization distribution.  
For a strongly unequal point double: 
\begin{equation}
 V_{1} \simeq -V_{2}(I_{1}/I_{2}) \label{eq:v_pred}
\end{equation}
\noindent where $V_2$ is the circular polarization on the weak component after 
self-calibration assuming zero circular polarization and $V_1$ is the extrapolated
circular polarization of the core.

The fractional error in this extrapolated measurement of $V_{1}$ is then the same as the 
fractional error in $V_{2}$ which is best determined by examination of the off-peak
noise in the image.  
In addition to this noise, our creation of model sources similar to 3C\,279 showed there
is a systematic loss of signal caused by the assumption of zero circular polarization 
on the order of $10$-$20$\%.  As a result, equation~\ref{eq:v_pred} will extrapolate
values for the core of 3C\,279 that are typically $10$-$20$\% low. 

For complicated sources, such as 3C\,84 and 3C\,273, it is much more difficult to
predict the effects of assuming no circular polarization during self-calibration.  
In the case of 3C\,84, we were fortunate because the real circular polarization
appeared in the jet, off the peak in the image.  As a result, {\em zero-V self-cal}
could not seriously reduce or relocate the circular polarization without inducing larger amounts
of (opposite sign) circular polarization on the map peak.  We verified this by carefully creating
detailed models of 3C\,84.  In doing so, we found that only the small negative component of
circular polarization in the {\em zero-V self-cal} image is (in part or in whole) an artifact 
of assuming no circular polarization during self-calibration.

\subsubsection{A Hybrid Technique: R/L Gain Transfer}
\label{s:gain-trans}

The second self-calibration assumption ($(RR+LL)/2 = \tilde{I}$) makes no 
assumption about circular polarization in the source and
accurately calibrates the average of R and L at a each antenna; however, it cannot correct the 
$R/L$ antenna gains crucial for detecting circular polarization.  The first
self-calibration assumption ($RR=\tilde{I}$, $LL=\tilde{I}$) can calibrate the $R/L$
antenna gains but may reduce or relocate real circular polarization in the process.

A natural compromise is to solve for the $R/L$ antenna gains only on objects which
are believed not to have circular polarization, using the assumption of zero 
circular polarization in self-calibration, and to transfer these measurements
to the other sources in the observation.

In practice, this is accomplished by first doing the complete calibration\footnote{Including 
removal of any feed leakage terms.} of the experiment with no assumption made 
about the presence or absence of circular polarization.  That is, 
the rigorous assumption $(RR+LL)/2 = \tilde{I}$ is made for all self-calibration steps.  
At the end of this calibration, only the $R/L$ antenna gain ratio remains to be
calibrated for each antenna.  Selected sources (which are assumed to contain no 
circular polarization) may then be self-calibrated with the assumption $RR=\tilde{I}$
and $LL=\tilde{I}$ to find the $R/L$ antenna gain ratios that will be applied (in some
averaged, smoothed form) to the remaining sources.

Initially, one does not know which sources have circular 
polarization and which do not.  For experiments with
a large number of sources that have well interleaved observations, it is possible to
initially assume that {\em all} sources have no circular polarization.  Self-calibration
on each source will then produce a series of $R/L$ gain corrections that are not 
directly applied to the sources but are merged, averaged, and smoothed on some
time interval long enough to span several source changes.  The resulting composite
correction table can then be applied directly to the data with little fear that real
circular polarization on any individual sources has strongly corrupted the results.
At this point, all the sources may be imaged in circular polarization.
If some sources appear to have significant circular polarization, we may go back a
step and construct a new gain table which omits the influence of those sources.

The technique described here was used to produce the {\em gain-transfer} images
presented in section~\ref{s:images}.  In the process of developing this technique we
tried a range of averaging times for the $R/L$ correction table from 4 hours to 
24 hours.  We found that, although the 4 hour averaging time produced images
with the least noise, our main results were essentially independent of averaging
time over this range.  We also found (with this range of averaging times) that  
we obtained essentially the same results when the circularly polarized sources
were allowed to contribute to the gain table, confirming that individual source
effects do not strongly influence the $R/L$ correction table if the averaging
time spans several source changes.

\begin{figure}
\begin{center}
\epsfig{file=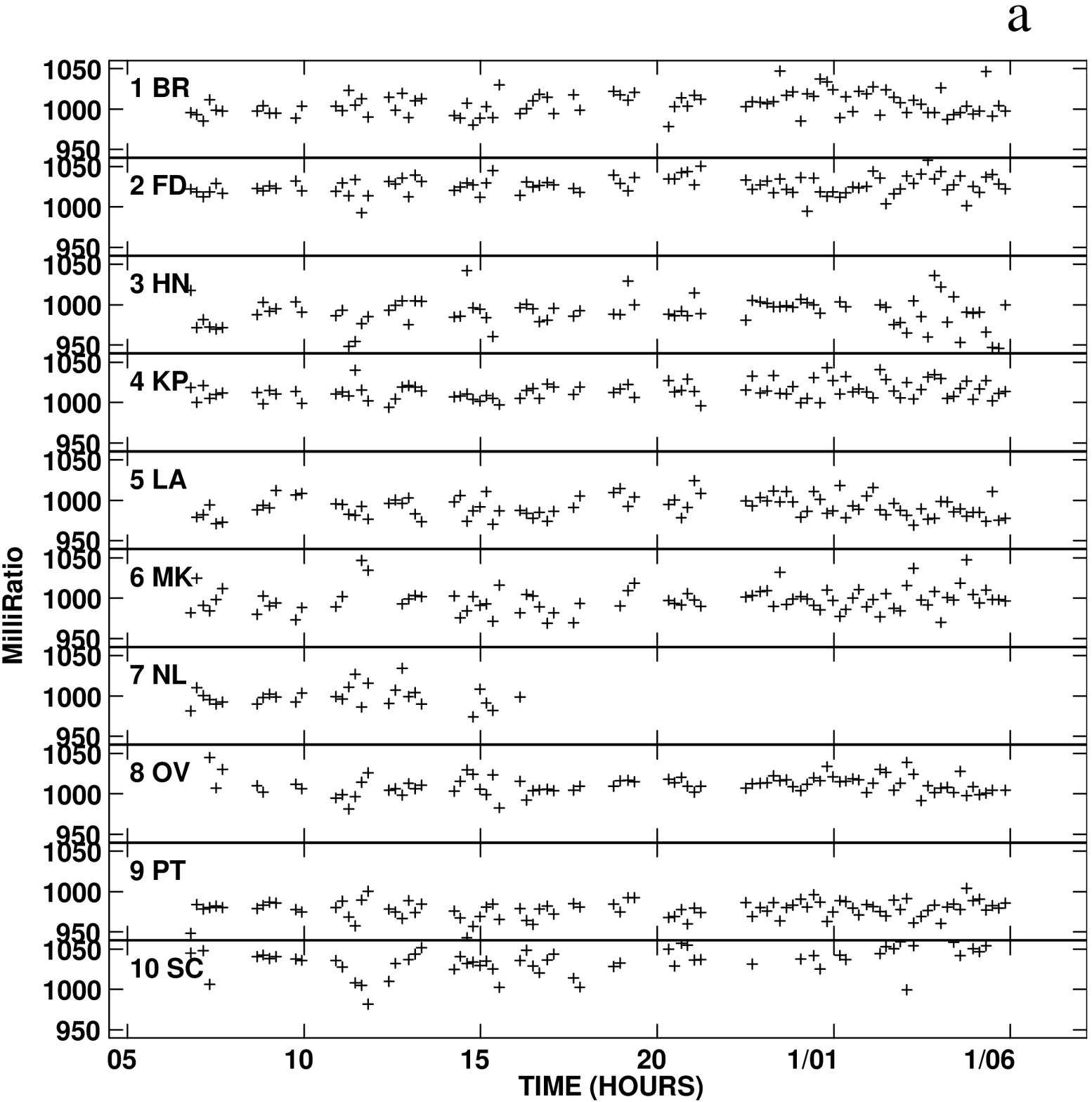,width=3.3in,angle=0} \\
\epsfig{file=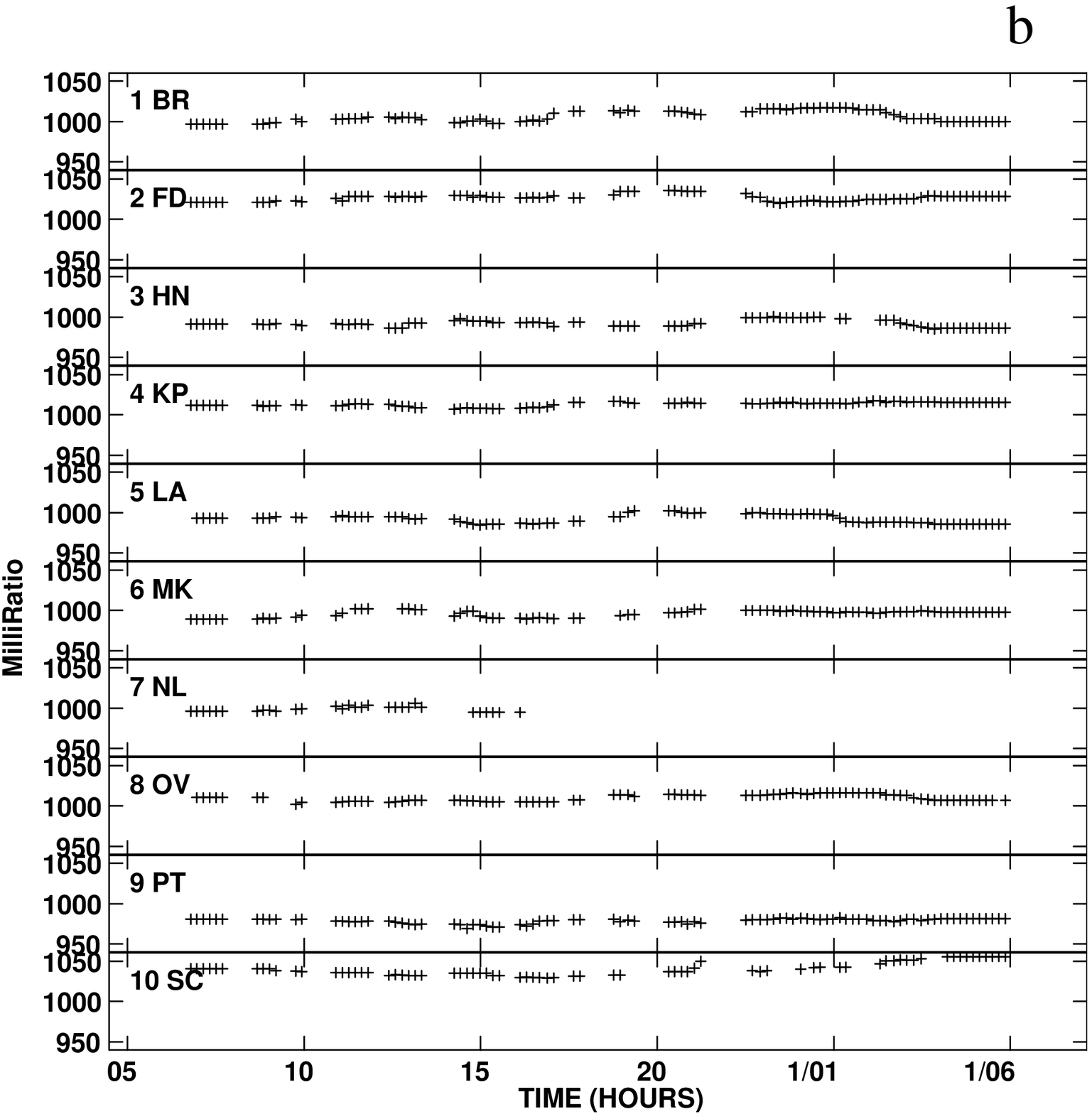,width=3.3in,angle=0} 
\end{center}
\caption{\label{f:gains}
$R/L$ amplitude gain ratio corrections for IF 2 of the 
1996.74 epoch.  Panel a) displays the raw corrections, before averaging and smoothing.  
Panel b) displays the corrections following averaging and smoothing on a 4 hour 
time scale.}
\end{figure}

\placefigure{f:gains}

Figure~\ref{f:gains} shows the $R/L$ amplitude gain ratios for IF 2 of the 
1996.74 epoch.  The raw and averaged corrections are both presented.  
The $R/L$ gain ratios have two time scales for variation.  
The first is a long time scale offset from unity on the order of a couple of percent.
This
offset varies slowly over the course of several hours up to 24 hours.  The second
time scale for variation is very short, on the order of a single scan.  These
short time scale variations typically have an rms deviation from the 24 hour
mean of $\lesssim$ 2\%.  Our procedure for calibrating the $R/L$ gain ratios corrects
the long time scale offset but cannot correct the very short time
scale fluctuations about this offset.  

We can estimate the effect of these short time scale variations by assuming
that the longer time scale offset has been correctly removed.  The
$R/L$ antenna gain ratios can then be parameterized by a small offset, $\delta$,
from unity.
\begin{equation}
(R/L)_i = 1+\delta_i \simeq \frac{1+\delta_i/2}{1-\delta_i/2}.
\end{equation}
\noindent So the $RR$ and $LL$ correlations are 
\begin{equation}
R_iR_j^* \simeq (1+\delta_i/2)(1+\delta_j/2)(\tilde{I}_{ij}+\tilde{V}_{ij})
\end{equation}
\noindent and
\begin{equation}
L_iL_j^* \simeq (1-\delta_i/2)(1-\delta_j/2)(\tilde{I}_{ij}-\tilde{V}_{ij})
\end{equation}
\noindent Then the measured $V$ visibilities become
\begin{equation}
\tilde{V}_{meas} \simeq \tilde{V}_{ij} + (\delta_i/2+\delta_j/2)\tilde{I}_{ij},
\end{equation}

Because the $\delta$ variations are essentially uncorrelated between antennas, scans,
and IFs, we can estimate the errors in our fractional circular polarization, $m_c$, 
measurements:
\begin{equation}
\label{e:gain-errs}
\Delta_{m_c} \simeq \frac{\delta}{\sqrt{N_aN_sN_{IF}}}
\end{equation}

\noindent where $N_a$ is the number of antennas, $N_s$ is the number of scans, and $N_{IF}$ is 
the number of IFs.  For our experiments $\delta \simeq 0.02$, $N_a = 10$, $N_s = 6$ to $10$,
and $N_{IF} = 4$, so we expect errors $\lesssim 0.15\%$ of the local total intensity in 
our circular polarization measurements by gain transfer.  This is consistent with the  
limits we see on sources which do not appear to have circular polarization.

\subsubsection{Phase-Only Mapping.}
\label{s:po2}

Wardle and Roberts\markcite{WR94} (1994) suggested mapping the closure 
phases of a VLBI data
set to detect a circularly polarized signal.  A very analogous (but simpler
to implement) idea is to map only the phases of the parallel hand correlations. 
Phase-only mapping involves setting all of the amplitudes of the parallel hand 
correlations to unity and using only the phase information to construct a
circular polarization map.  This method is useful because errors in the antenna
amplitude gains dominate the error in the other techniques for making 
circular polarization images.  By ignoring the amplitudes, we look for the signal 
directly in the phases of the data.

For simple sources, phase-only images
will be anti-symmetric to first order and the true location of the 
circular polarization signal is ambiguous.  For example, consider the case of a
point double (like 3C\,279) with the stronger peak, $I_1$, taken to be at the 
map center and the weaker peak, $I_2$, at position ($x_2$, $y_2$).  A small 
amount of circular polarization, $V_0$, is located on either of the two peaks.

If $V_0$ coincides with the position of $I_2$ (the weaker component), the
phase-only map is given by
\begin{equation}
V_P(x,y) \approx \frac{V_0}{2I_1}[\delta (x-x_2,y-y_2) - \delta (x+x_2,y+y_2)].
\end{equation}

If $V_0$ coincides with the position of $I_1$ (the stronger component), the phase-only
map is given by
\begin{equation}
\label{e:phase-only}
V_P(x,y) \approx -\frac{V_0I_2}{2I_1^2}[\delta (x-x_2,y-y_2) - \delta (x+x_2,y+y_2)].
\end{equation}

These two phase-only maps differ only in the strength of the  
components.  Without a-priori knowledge of the strength or true location
of $V_0$, it is not possible to distinguish between them.  However, if we
do know the true location of $V_0$, its strength can be deduced using
these expressions.  

It is interesting to note that relative amplitude information is preserved
in phase-only mapping, although the phase-only images are independent of the 
antenna amplitude gains. This seeming contradiction is due to the fact that 
while we have forced the $RR$ and $LL$ correlations to have unit amplitude 
(thus eliminating any dependence on antenna amplitude gains), it is the
{\em total} amplitude of these correlations which have been set to unity.
The amplitudes of individual $I$ and $V$ components which conspire to 
form $RR$ and $LL$ remain in the form of ratios if the source structure 
is more complicated than a point.

\subsection{Beam Squint}

Beam squint is a problem for any antenna with off-axis feed elements.  The 
left and right hand feeds have slightly displaced
primary beams.  Any pointing error at the telescope will cause
one feed to receive a larger signal than the other, resulting in
a small, artificial amplitude difference between the RCP and LCP signals
 at that antenna. 

For the VLBA at 22 GHz, a 7'' pointing error (typical for the VLBA,\markcite{N95}
Napier 1995)
will result in a 1\% amplitude difference between RCP and LCP signals at an
antenna\markcite{WPC} (Walker, personal communication).  The observations 
presented in this paper are at 15 GHz
where the problem is $(1.5)^2$ smaller due to the larger primary beams of
each feed. The placement of the 15 GHz feed in VLBA antennas is also
better than the 22 GHz feed as it is not parallel to the azimuthal axis
where the largest pointing errors occur\markcite{WPC} 
(Walker, personal communication).

Fortunately beam squint will not be correlated between antennas and
should vary with azimuthal antenna rotation during the observations.
Any beam squint should be a pure amplitude gain error at
each antenna and will be completely removed by the {\em zero-V self-cal}
technique discussed in~\ref{s:zerov}. In the {\em gain transfer} technique,
beam squint effects should show up as short term $R/L$ amplitude 
variations and, as such, are already included in our error estimates 
for those images. Beam squint will not
affect the phases of the data, and therefore images produced by {\em
phase-only} mapping will be independent of beam squint effects.
 
\subsection{Baseline Based Errors}

Real circular polarization would be completely calibrated out of a data set
by {\em baseline based} self-calibration which assumes $RR =
\tilde{I}_{mod}$ and $LL = \tilde{I}_{mod}$.  This makes it difficult
to remove any residual baseline based errors from the data if we wish to
detect circular polarization. 

However, 
there is excellent evidence that the baseline based errors on the VLBA are
extraordinarily small.  Imaging of the nearly unresolved point source,
DA\,193, at 5 GHz with a dynamic range of better than 100,000 to 1 indicates that
the baseline based errors are no larger than 0.1\% at that frequency
\markcite{W95} (Walker 1995).  
In our simulations 
we added baseline based errors much larger than this to our model
data sets and found that while they increased the noise in the final
circular polarization images they did not generate a spurious 
circularly polarized signal.

\subsection{Additional Calibration Tests}

The calibration checks described in this section were conducted early-on 
in our investigation.  The results increased our confidence in our ability
to calibrate for circular polarization and served as a basis for the
more systematic tests and investigations described in the previous sections.
We briefly present these initial tests and results here for completeness and
to provide other observers with suggestions for other ``common sense'' checks
when calibrating for circular polarization.

For epoch 1996.57, we repeated the entire calibration procedure 
(post fringe-fit) on the source 3C\,84 for each IF channel
separately, and then omitting each antenna in turn from the array. In no
case did the apparent circularly polarized signal change significantly. It
was present in every IF channel, and could not be attributed to the behavior
of any one antenna.

We also checked the effect of incorrect subtraction of antenna feed leakage
(D-terms) from the parallel hand data for the 1996.57 epoch of 3C\,279. 
This was done by simply omitting the correction altogether and also by applying
a completely incorrect set of D-terms.  Both procedures increased the 
noise on the {\em zero-V self-cal} image enough that one might misinterpret 
the real circular polarization of the source as noise.  Correct application of the
D-term solutions, however, revealed a clean (real) circular polarization signal in the
{\em zero-V self-cal} image.  

We conducted a similar test on 3C\,84.  In this case, the {\em zero-V self-cal}
image created from data without the D-terms removed was nearly the same as the
the {\em zero-V self-cal} image created from data where we performed the 
proper D-term corrections.  This is exactly what we expected for 3C\,84 which has
no linear polarization to leak into the circular polarization. 
The total intensity can still leak into the circular polarization image, but 
it is a much smaller effect.

Finally, we created three model sources, similar to 3C\,279 in
total intensity and linear polarization structure, with either no initial $V$, $V$ on
the core, or $V$ on the jet component.  These models contained thermal noise,
D-term leakage, time-dependent complex gains, and baseline-based errors.  We
carried these models through our standard calibration procedures
and found that we were able to detect a circularly polarized signal placed in 
the data.  Equally important, we were unable to detect a circularly polarized signal 
when the model contained none.



\newpage

\clearpage


   







\input{homan.tab1.tex}

\input{homan.tab2.tex}

\input{homan.tab3.tex}

\input{homan.tab4.tex}

\input{homan.tab5.tex}

\input{homan.tab6.tex}

\input{homan.tab7.tex}
\input{homan.tab8.tex}

\end{document}

%% file: homan.tab1.tex
\begin{table}
\begin{scriptsize}
\begin{center}
\tablenum{1}
\caption[]{\label{t:epochs} Epochs of observation.\\} 
\begin{tabular}{ccc}
\tableline \tableline 
$Date$ & $Epoch$ & $Notes$ \\ 
\tableline 
Jan 19 & $1996.05$ & \tablenotemark{a}  \\
Mar 22 & $1996.23$ & \tablenotemark{b}  \\
May 27 & $1996.41$ & \tablenotemark{c}  \\
Jul 27 & $1996.57$ &   \\
Sep 27 & $1996.74$ & \tablenotemark{d} \quad \tablenotemark{e}  \\
\tableline \tableline
\end{tabular} 
\tablenotetext{a}{North Liberty antenna off-line for entire observation.} 
\tablenotetext{b}{Owens Valley antenna off-line for first half of observation.} 
\tablenotetext{c}{No fringes found to the Kitt Peak antenna.} 
\tablenotetext{d}{North Liberty antenna off-line for second half of observation.}  
\tablenotetext{e}{Some data loss from the Owens Valley antenna.} 
\end{center} 
\end{scriptsize}       
\end{table}

%% file: homan.tab2.tex
\begin{table}
\begin{scriptsize}
\begin{center}
\tablenum{2}
\caption[]{\label{t:sources} Source names and redshifts.\\} 
\begin{tabular}{cccc}
\tableline \tableline 
$J2000$ & $B1950$ & $Other$ & $Redshift$ \\ 
\tableline 
$J0319+41$ & $0316+413$ & 3C\,84 & $0.018$ \\
$J0433+05$ & $0430+052$ & 3C\,120 & $0.033$ \\ 
$J0530+13$ & $0528+134$ & PKS 0528$+$134 & $2.070$ \\
$J0738+17$ & $0735+178$ & PKS 0735$+$178 & $0.424$\tablenotemark{a} \\
$J0854+20$ & $0851+202$ & OJ287 & $0.306$ \\
$J1224+21$ & $1222+216$ &       & $0.435$ \\
$J1229+02$ & $1226+023$ & 3C\,273 & $0.158$ \\
$J1256-05$ & $1253-055$ & 3C\,279 & $0.536$ \\
$J1310+32$ & $1308+326$ &         & $0.996$ \\
$J1512-09$ & $1510-089$ &         & $0.360$ \\
$J1751+09$ & $1749+096$ & 4C\,09.56 & $0.322$ \\
$J1927+73$ & $1928+738$ & 4C\,73.18 & $0.302$ \\
$J2005+77$ & $2007+776$ &           & $0.342$ \\ 
\tableline \tableline 
\end{tabular} 
\tablenotetext{a}{Absorption line redshift.} 
\end{center} 
\end{scriptsize} 
\end{table}

%% file: homan.tab3.tex
\begin{table}
\begin{scriptsize}
\begin{center}
\tablenum{3}
\caption[]{\label{t:3c279} Core flux and polarization of 3C\,279.\\}
\begin{tabular}{cccccccccc} 
\tableline \tableline
$Epoch$ &
$Component$ & $R$ & $\Theta$ & $I$ &
$P$ & $\chi$ & $m_L$ & $V$ & $m_C$ \\
&& (mas) & (deg) & (Jy) & (Jy) & (deg) & (\%) & (Jy) & (\%) \\
\tableline
1996.05 & CE & $\ldots$ & $\ldots$ & $8.39$ & $0.283$ & $91$ & $3.4$  & $\ldots$ & $\ldots$ \\
        & CW & $0.15$   & $-117$   & $3.12$ & $0.672$ & $19$ & $21.5$ & $+0.036$ ($\pm 0.018$) & $+1.2$ \\
&&&&&&&&& \\
1996.23 & CE & $\ldots$ & $\ldots$ & $8.53$ & $0.379$ & $-80$ & $4.4$  & $\ldots$ & $\ldots$ \\
        & CW & $0.11$   & $-126$   & $5.33$ & $0.591$ & $24$  & $11.1$  & $+0.099$ ($\pm 0.018$) & $+1.9$ \\
&&&&&&&&& \\
1996.41 & CE & $\ldots$ & $\ldots$ & $9.23$ & $1.01$  & $-78$ & $11.0$ & $\ldots$ & $\ldots$ \\
        & CW & $0.10$   & $-115$   & $6.80$ & $0.944$ & $-167$ & $13.9$ & $+0.092$ ($\pm 0.020$) & $+1.4$ \\
&&&&&&&&& \\
1996.57 & CE & $\ldots$ & $\ldots$ & $7.58$ & $0.814$ & $-83$ & $10.7$ & $\ldots$ & $\ldots$ \\
        & CW & $0.09$   & $-115$   & $9.19$ & $0.914$ & $-161$ & $9.9$ & $+0.089$ ($\pm 0.020$) & $+1.0$ \\
&&&&&&&&& \\
1996.74 & CE & $\ldots$ & $\ldots$ & $7.08$ & $0.825$ & $-106$ & $11.6$ & $\ldots$ & $\ldots$ \\
        & CW & $0.12$   & $-125$   & $9.90$ & $0.986$ & $6$    & $10.0$ & $+0.070$ ($\pm 0.020$) & $+0.7$ \\
\tableline \tableline
\end{tabular}
\tablecomments{Circular polarization is assumed to be associated with the CW component
of the core.  The circular polarization measurements are from the {\em gain transfer} calibration
technique.  The off-peak RMS noise in the circular polarization images was typically 1-2 mJy/beam;
however, the errors in our measurements are limited by the short time-scale $R/L$ gain 
fluctuations to $\lesssim 0.15\%$ of the total $I$ of the core.}
\end{center}
\end{scriptsize}
\end{table}

%% file: homan.tab4.tex
\begin{table}
\begin{scriptsize}
\begin{center}
\tablenum{4}
\caption[]{\label{t:3c279-alt} Raw zero-V and phase-only results for 3C\,279.\\}
\begin{tabular}{cccc} 
\tableline \tableline
\multicolumn{1}{c}{} &
\multicolumn{1}{c}{Zero-V Self-Cal} &
\multicolumn{2}{c}{Phase-Only Map} \\
Epoch & $V_{jet}$ & $V_{jet}$ & $V_{anti-jet}$ \\
& (mJy) & (mJy) & (mJy) \\
\tableline
1996.05 & $-5.6$ & $-0.23$ & $0.20$ \\    
1996.23 & $-7.6$ & $-0.34$ & $0.32$ \\    
1996.41 & $-5.4$ & $-0.20$ & $0.19$ \\    
1996.57 & $-11.3$ & $-0.37$ & $0.34$ \\   
1996.74 & $-11.6$ & $-0.36$ & $0.35$ \\   
\tableline \tableline
\end{tabular}
\tablecomments{The values were obtained from model fitting with point
sources in the u-v plane.  The off-peak RMS noise in the {\em zero-V self-cal}
images was typically $0.5$ mJy/beam.  The off-peak RMS noise in the {\em phase-only}
images was typically $0.02$ mJy/beam.}
\end{center}
\end{scriptsize}
\end{table}

%% file: homan.tab5.tex
\begin{table}
\begin{scriptsize}
\begin{center}
\tablenum{5}
\caption[]{\label{t:3c279-2pt-ext} Extrapolated CP using a simplified point double model for 3C\,279.\\}
\begin{tabular}{ccccccccc} 
\tableline \tableline
\multicolumn{1}{c}{} &
\multicolumn{2}{c}{I Model} &
\multicolumn{2}{c}{Zero-V Self-Cal} &
\multicolumn{2}{c}{Phase-Only Map} &
\multicolumn{2}{c}{Gain Transfer} \\
Epoch & $I_{core}$ & $I_{jet}$ & 
\multicolumn{2}{c}{Extrapolated $V_{core}$} &  
\multicolumn{2}{c}{Extrapolated $V_{core}$} &  
\multicolumn{2}{c}{$V_{core}$} \\  
& (Jy) & (Jy) & 
\multicolumn{2}{c}{(mJy)} &
\multicolumn{2}{c}{(mJy)} &
\multicolumn{2}{c}{(mJy)} \\
\tableline
1996.05 & $11.40$ & $2.02$ & \multicolumn{2}{c}{$32$~($+9$,$-3$)} 
                           & \multicolumn{2}{c}{$27$~($+7$,$-2$)} 
                           & \multicolumn{2}{c}{$36$~$\pm 18$}  \\
1996.23 & $13.14$ & $2.04$ & \multicolumn{2}{c}{$49$~($+14$,$-4$)} 
                           & \multicolumn{2}{c}{$65$~($+19$,$-4$)} 
                           & \multicolumn{2}{c}{$99$~$\pm 18$}  \\
1996.41 & $15.32$ & $1.84$ & \multicolumn{2}{c}{$45$~($+14$,$-4$)} 
                           & \multicolumn{2}{c}{$49$~($+17$,$-5$)} 
                           & \multicolumn{2}{c}{$92$~$\pm 20$}  \\
1996.57 & $15.58$ & $2.14$ & \multicolumn{2}{c}{$82$~($+20$,$-4$)} 
                           & \multicolumn{2}{c}{$81$~($+20$,$-4$)} 
                           & \multicolumn{2}{c}{$89$~$\pm 20$} \\
1996.74 & $15.57$ & $2.32$ & \multicolumn{2}{c}{$78$~($+20$,$-4$)} 
                           & \multicolumn{2}{c}{$75$~($+19$,$-4$)} 
                           & \multicolumn{2}{c}{$70$~$\pm 20$} \\
\tableline \tableline
\end{tabular}
\tablecomments{Errors on these extrapolations include a positive offset
for the possibility that self-calibration assuming no circular polarization may incur
an overall signal loss of up to about $20\%$ (see \S{A.2.1}).  
Measurements of the circular polarization from {\em gain transfer} are included 
for comparison.}
\end{center}
\end{scriptsize}
\end{table}

%% file: homan.tab6.tex
\begin{table}
\begin{scriptsize}
\begin{center}
\tablenum{6}
\caption[]{\label{t:J0530}
Core flux and polarization measurements for PKS 0528+134.\\}  
\begin{tabular}{cccccccccc} 
\tableline \tableline
$Epoch$ &
$Component$ & $R$ & $\Theta$ & $I$ &
$P$ & $\chi$ & $m_L$ & $V$ & $m_C$ \\
&& (mas) & (deg) & (Jy) & (Jy) & (deg) & (\%) & (Jy) & (\%) \\
\tableline
1996.05 & C & $\ldots$ & $\ldots$ & $8.65$ & $0.207$ & $100$ & $2.4$  & $+0.044$ ($\pm 0.014$) & $+0.5$ \\
&&&&&&&&& \\
1996.23 & C & $\ldots$ & $\ldots$ & $9.19$ & $0.053$ & $52$ & $0.6$  & $+0.107$ ($\pm 0.012$) & $+1.2$ \\
&&&&&&&&& \\
1996.41 & C & $\ldots$ & $\ldots$ & $7.96$ & $0.072$  & $-42$ & $0.9$ & $+0.040$ ($\pm 0.010$) & $+0.5$ \\
&&&&&&&&& \\
1996.57 & C & $\ldots$ & $\ldots$ & $7.76$ & $0.137$ & $-90$ & $1.8$ & $+0.045$ ($\pm 0.010$) & $+0.6$ \\
&&&&&&&&& \\
1996.74 & C & $\ldots$ & $\ldots$ & $8.19$ & $0.246$ & $-97$ & $2.7$ & $+0.047$ ($\pm 0.011$) & $+0.6$ \\
\tableline \tableline
\end{tabular}
\tablecomments{
The circular polarization measurements are from the {\em gain transfer} calibration
technique.  The off-peak RMS noise in the circular polarization images was typically 1 mJy/beam;
however, the errors in our measurements are limited by the short time-scale $R/L$ gain 
fluctuations to $\lesssim 0.15\%$ of the $I$ peak.}
\end{center}
\end{scriptsize}
\end{table}

%% file: homan.tab7.tex
\begin{table}
\begin{scriptsize}
\begin{center}
\tablenum{7}
\caption[]{\label{t:3c273}
Core flux and polarization measurements for 3C\,273.\\}  
\begin{tabular}{cccccccccc} 
\tableline \tableline
$Epoch$ &
$Component$ & $R$ & $\Theta$ & $I$ &
$P$ & $\chi$ & $m_L$ & $V$ & $m_C$ \\
&& (mas) & (deg) & (Jy) & (Jy) & (deg) & (\%) & (Jy) & (\%) \\
\tableline
1996.05 & C & $\ldots$ & $\ldots$ & $2.81$ & $< 0.010$ & $\ldots$ & $< 0.3$  & $\sim -0.010$ ($\pm 0.010$) & $\sim -0.4$ \\
&&&&&&&&& \\
1996.23 & C & $\ldots$ & $\ldots$ & $6.42$ & $< 0.010$ & $\ldots$ & $< 0.2$  & $< 0.010$ ($\ldots$) & $< 0.2$ \\
&&&&&&&&& \\
1996.41 & C & $\ldots$ & $\ldots$ & $11.55$ & $0.034$  & $-178$ & $0.3$ & $-0.074$ ($\pm 0.015$) & $-0.6$ \\
&&&&&&&&& \\
1996.57 & C & $\ldots$ & $\ldots$ & $13.38$ & $0.076$ & $-149$ & $0.6$ & $-0.072$ ($\pm 0.017$) & $-0.5$ \\
&&&&&&&&& \\
1996.74 & CE & $\ldots$ & $\ldots$ & $6.51$ & $0.035$ & $-78$ & $0.5$ & $-0.019$ ($\pm 0.008$) & $-0.3$ \\
& CW & $0.48$ & $-120$ & $7.81$ & $0.039$ & $0$ & $0.5$ & $-0.037$ ($\pm 0.010$) & $-0.5$ \\
\tableline \tableline
\end{tabular}
\tablecomments{
The circular polarization measurements are from the {\em gain transfer} calibration
technique.  The off-peak RMS noise in the circular polarization images was typically 1-2 mJy/beam;
however, the errors in our measurements are limited by the short time-scale $R/L$ gain 
fluctuations to $\lesssim 0.15\%$ of the $I$ peak.  The {\em gain transfer} circular
polarization result from the 1995.05 epoch is particularly noisy and has a larger estimated error.}
\end{center}
\end{scriptsize}
\end{table}

%% file: homan.tab8.tex
\begin{table}
\begin{scriptsize}
\begin{center}
\tablenum{8}
\caption[]{\label{t:rest} Core circular polarization.\\} 
\begin{tabular}{ccccccc}
\tableline \tableline 
$Source$ & $Epoch$ & $m_c$ & $+/-$ &
$I$ & $m_L$ & $\chi$ \\ 
&& (\%) & (\%) & (Jy) & (\%) & (deg) \\
\tableline 
3C\,120 & 1996.05 & $< 0.3$ & $\ldots$ & $1.20$ & $0.2$ & $-9$  \\
        & 1996.23 & $< 0.2$ & $\ldots$ & $1.39$ & $0.4$ & $-88$ \\
         & 1996.41 & $< 0.3$ & $\ldots$ & $1.22$ & $0.3$ & $70$ \\
         & 1996.57 & $< 0.2$ & $\ldots$ & $0.85$ & $0.0$ & $\ldots$ \\
         & 1996.74 & $< 0.2$ & $\ldots$ & $1.47$ & $0.1$ & $-83$ \\
&&&&&& \\
J0738+17 & 1996.05 & $\sim+0.4$ & $0.4$ & $0.62$ & $3.1$ & $-61$ \\
         & 1996.23 & $< 0.4$ & $\ldots$ & $0.52$ & $1.3$ & $-54$  \\
         & 1996.41 & $< 0.3$ & $\ldots$ & $0.52$ & $1.1$ & $70$ \\
         & 1996.57 & $< 0.4$ & $\ldots$ & $0.46$ & $1.0$ & $-77$  \\
         & 1996.74 & $< 0.4$ & $\ldots$ & $0.49$ & $1.7$ & $-45$  \\
&&&&&& \\
OJ287 & 1996.05 & $< 0.1$ & $\ldots$ & $2.11$ & $3.9$ & $8$    \\
         & 1996.23 & $< 0.2$ & $\ldots$ & $1.51$ & $2.8$ & $-8$   \\
         & 1996.41 & $< 0.2$ & $\ldots$ & $1.45$ & $2.8$ & $-2$   \\
         & 1996.57 & $< 0.2$ & $\ldots$ & $1.06$ & $2.8$ & $-15$  \\
         & 1996.74 & $< 0.2$ & $\ldots$ & $1.49$ & $1.9$ & $-55$  \\
&&&&&& \\         
J1224+21 & 1996.23 & $< 0.2$ & $\ldots$ & $1.42$ & $5.0$ & $-49$  \\
         & 1996.41 & $< 0.1$ & $\ldots$ & $1.16$ & $5.5$ & $-61$  \\
         & 1996.57 & $< 0.2$ & $\ldots$ & $1.16$ & $3.9$ & $-52$  \\
         & 1996.74 & $< 0.1$ & $\ldots$ & $1.34$ & $2.5$ & $-49$  \\
&&&&&& \\
J1310+32 & 1996.05 & $< 0.1$ & $\ldots$ & $2.81$ & $3.0$ & $26$  \\
         & 1996.23 & $< 0.1$ & $\ldots$ & $2.68$ & $1.2$ & $29$  \\
         & 1996.41 & $< 0.1$ & $\ldots$ & $2.42$ & $0.3$ & $43$ \\
         & 1996.57 & $< 0.1$ & $\ldots$ & $2.30$ & $2.6$ & $6$ \\
         & 1996.74 & $< 0.2$ & $\ldots$ & $2.33$ & $2.2$ & $4$    \\
\tableline \tableline 
\end{tabular} 
\end{center} 
\end{scriptsize} 
\end{table} 

\begin{table}
\begin{scriptsize}
\begin{center}
\begin{tabular}{ccccccc}
\tableline \tableline 
$Source$ & $Epoch$ & $m_c$ & $+/-$ &
$I$ & $m_L$ & $\chi$ \\ 
&& (\%) & (\%) & (Jy) & (\%) & (deg) \\
\tableline 
J1512-09 & 1996.05 & $< 0.2$ & $\ldots$ & $0.96$ & $1.1$ & $7$    \\
         & 1996.23 & $\sim+0.2$ & $0.2$ & $1.32$ & $1.6$ & $10$  \\
         & 1996.41 & $< 0.2$ & $\ldots$ & $1.54$ & $1.7$ & $86$  \\
         & 1996.57 & $< 0.1$ & $\ldots$ & $1.49$ & $2.3$ & $30$  \\
         & 1996.74 & $\sim+0.2$ & $0.2$ & $1.18$ & $2.2$ & $-44$ \\
&&&&&& \\ 
J1751+09 & 1996.05 & $\sim-0.1$ & $0.1$ & $2.73$ & $1.6$ & $-71$  \\
         & 1996.23 & $< 0.2$ & $\ldots$ & $1.04$ & $3.8$ & $-71$   \\
         & 1996.41 & $< 0.2$ & $\ldots$ & $0.78$ & $0.6$ & $74$  \\
         & 1996.57 & $< 0.2$ & $\ldots$ & $0.79$ & $1.4$ & $-74$   \\
         & 1996.74 & $< 0.2$ & $\ldots$ & $0.86$ & $1.1$ & $-47$   \\
&&&&&& \\
J1927+73 & 1996.05 & $\sim+0.3$ & $0.1$ & $2.07$ & $0.5$ & $-82$  \\
         & 1996.23 & $\sim+0.2$ & $0.1$ & $2.17$ & $0.6$ & $-32$ \\
         & 1996.41 & $< 0.1$ & $\ldots$ & $2.25$ & $1.2$ & $-56$  \\
         & 1996.57 & $< 0.2$ & $\ldots$ & $2.31$ & $0.6$ & $-38$  \\
         & 1996.74 & $\sim+0.3$ & $0.1$ & $2.52$ & $0.6$ & $-18$ \\
&&&&&& \\                   
J2005+77 & 1996.05 & $< 0.4$ & $\ldots$ & $0.70$ & $8.4$ & $79$ \\
         & 1996.23 & $\sim-0.4$ & $0.4$ & $0.63$ & $6.3$ & $79$ \\
         & 1996.41 & $< 0.2$ & $\ldots$ & $0.55$ & $4.1$ & $-93$ \\
         & 1996.57 & $< 0.4$ & $\ldots$ & $0.57$ & $3.8$ & $-84$ \\
         & 1996.74 & $< 0.3$ & $\ldots$ & $0.65$ & $4.3$ & $-80$ \\
\tableline \tableline 
\end{tabular} 
\tablecomments{Measurements were made from
{\em gain-transfer} images.  The upper limits and errors are estimated 
to be roughly $\sqrt{2}$ times the peak noise in the images.}
\end{center} 
\end{scriptsize} 
\end{table}